\pdfoutput=1
\documentclass[12 pt]{article}
\usepackage[utf8]{inputenc}
\usepackage[top=1.5cm, bottom=2cm, left=2.5cm, right=2cm]{geometry}
\usepackage{amsmath}
\usepackage{physics}
\usepackage{graphicx}
\usepackage{subcaption}
\usepackage{cite}

\def\e{{\rm e}}

\newcommand{\be}{\begin{equation}}
\newcommand{\ee}{\end{equation}}
\newcommand{\ba}{\begin{align}}
\newcommand{\ea}{\end{align}}
\newcommand{\bg}{\begin{gather}}
\newcommand{\eg}{\end{gather}}
\newcommand{\bseq}{\begin{subequations}}
\newcommand{\eseq}{\end{subequations}}

\makeatletter

\def\lsim{\compoundrel<\over\sim}
\def\compoundrel#1\over#2{\mathpalette\compoundreL{{#1}\over{#2}}}
\def\compoundreL#1#2{\compoundREL#1#2}
\def\compoundREL#1#2\over#3{\mathrel
         {\vcenter{\hbox{$\m@th\buildrel{#1#2}\over{#1#3}$}}}}
\makeatother

\begin{document}
\newcommand{\Lagr}{\mathcal{L}}

\title{Lunar neutrinos}
\author{S. Demidov${}^{1,2}$, D. Gorbunov${}^{1,2}$\\
${}^1$ 
  Institute for Nuclear Research of the Russian Academy of Sciences, \\
Moscow 117312, Russia\\
${}^2$ Moscow Institute of Physics and Technology, \\
Dolgoprudny 141700, Russia\\ 
}

\date{}

\maketitle

\begin{abstract}
 Cosmic rays bombard the lunar surface producing mesons, which
 attenuate inside the regolith. They get slower and decay weakly
 into mostly sub-GeV neutrinos leaving the surface. Thus the Moon
 shines in neutrinos. Here we calculate spectra of low energy neutrinos,
 which exhibit bright features potentially recognisable above 
 isotropic neutrino background in the direction towards the Moon. Their observation, though
 a very challenging task for future neutrino large volume experiments,
 would make the Moon the nearest astrophysical source for which
 the concept of multimessenger astronomy works truly. Remarkably,
 some features of the lunar neutrino flux are sensitive to the
 surface mass density of the Moon.  
\end{abstract}

\section{Introduction}

So far we have observed only a single extra-terrestrial permanent
neutrino source, that is the Sun. Neutrinos can emerge upon hadron
production in astrophysical sources,
see e.g.\,\cite{Anchordoqui:2013dnh,Vitagliano:2019yzm}, 
however the powerful ones are at quite large
distances, so the individual positional identification requires
an angular resolution far beyond the capabilities of the operating neutrino
telescopes. Naturally, on this way it is tempting to start with a close
source with obvious production mechanism to have both theoretical
prediction robust and positional identification reliable.

A promising source of this kind is the Moon. Cosmic rays (CR) hit its
surface freely, provided the absence of atmosphere. They initiate
hadronic cascades developing inside the regolith, and numerous
mesons weakly decay producing neutrinos. The
cosmic ray spectrum degrades with energy and the scattering off
regolith slows down the mesons, so the largest neutrino flux is expected from
stopped mesons and muons. Hence the Moon becomes a source of neutrinos
in sub-GeV energy range to be observed on the sky by neutrino telescopes.  
Semianalytic calculations of high energy (above 10\,GeV)
neutrino energy part of the spectrum were performed in the study~\cite{Miller:2006cn}
(see also~\cite{1965ICRC....2.1039V,Volkova1989} for earlier studies).
It showed a suppression of the lunar neutrino flux as compared
to the atmospheric one calculated over the Moon's solid angle by factor
of $10^{-2}-10^{-4}$ depending on neutrino energy. Remarkably, both lunar
and atmospheric neutrinos have the same origin in cosmic rays. 

In this letter we calculate the low energy part (from 10~MeV to about
10~GeV) of lunar neutrino spectrum and compare the lunar neutrino
flux at the Earth with isotropic background from atmospheric
  and supernova neutrinos. We numerically
simulate interactions of cosmic rays with the regolith and count
neutrinos produced in meson and muon decays. We observe that although
the ratio of total lunar to atmospheric neutrino fluxes in the low
energy range is close 
to unity (contrary to the more energetic part) their spectra are
quite different. This observation makes it potentially possible
to distinguish between neutrinos of different origin, although
the small Moon's solid angle with respect to
  the whole sky ($\sim 5\times 10^{-6}$) makes the task of lunar neutrino
  detection a great challenge even for near future neutrino experiments
  at the Earth.

Note that previous studies of the impact of cosmic ray bombarding the Moon
include in particular simulations of the production of cosmogenic
nuclides~\cite{doi:10.1002/2016JA023308} and gamma-ray
albedo~\cite{Moskalenko:2007mk}. Induced by the cosmic rays, 
$\gamma$-ray emission from the Moon was measured by
FERMI~\cite{Abdo:2012nfa,Cerutti:2016gts}. A Moon shadow in high energy
cosmic rays was measured in numerous experiments, most recently
by ANTARES~\cite{Distefano:2011zza,Albert:2018yoj} and
IceCube~\cite{Boersma:2010zz,Aartsen:2013zka}.
Possible implications of interactions of astrophysical neutrinos in the
Moon were discussed in~\cite{Fargion:2017lok}.

\section{Simulation details and results}

Absence of atmosphere makes the Moon a very effective cosmic ray
dump. Namely, dominant sources of neutrinos, i.e. pions and kaons,
which 
emerged in collisions of cosmic rays with the Moon soil, stop before decay,
hence producing monoenergetic neutrinos.

\begin{figure}[!htb]
  \begin{center}
  \includegraphics[width=11.0cm]{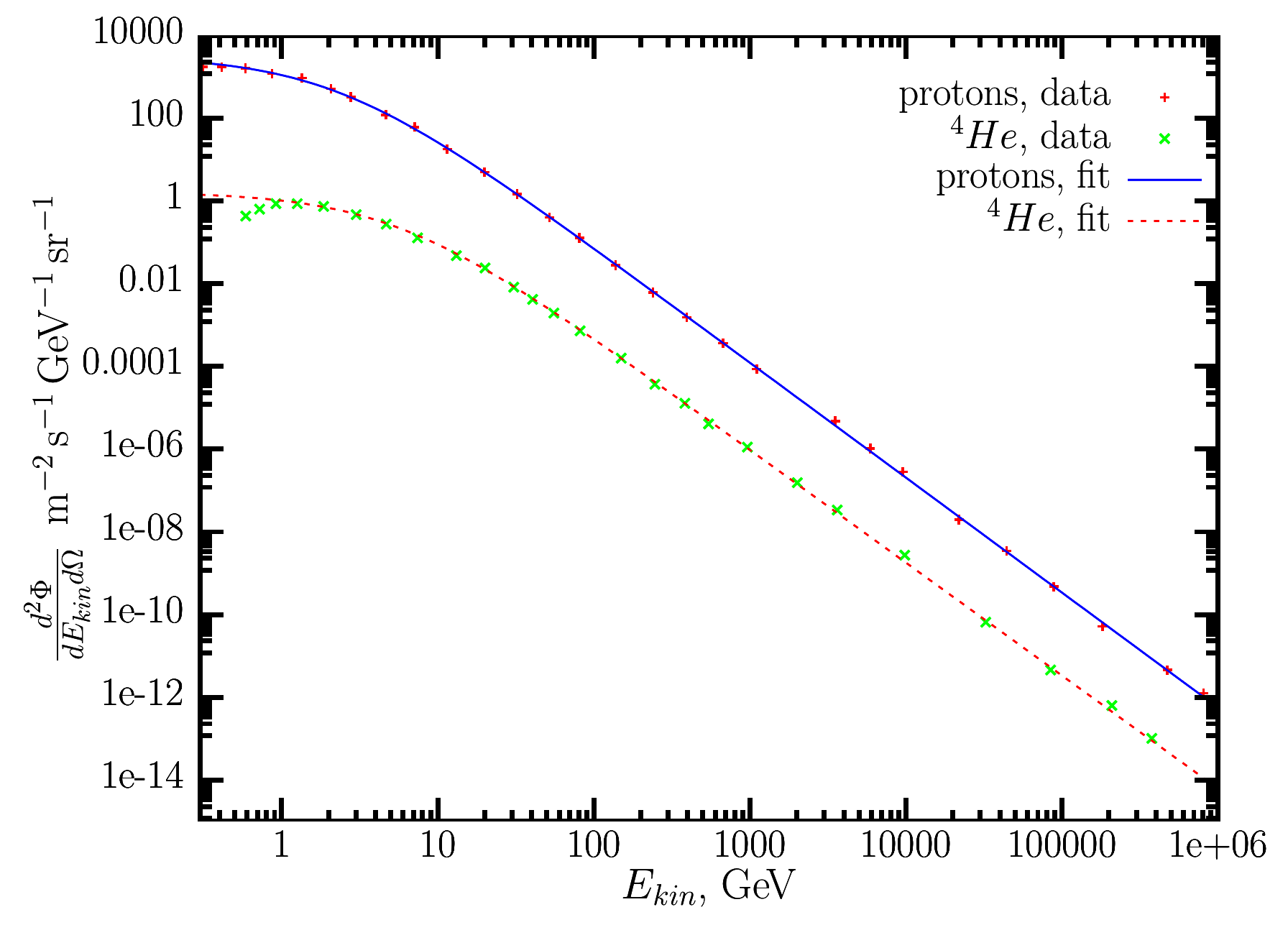}
  \caption{\label{fig1} Proton and $^4$He measured fluxes (the latter is scaled by
    $10^{-2}$ factor as in PDG) and numerical fits to them used in our analysis.}  
  \end{center}
\end{figure}

To simulate neutrino flux from the Moon we exploit GEANT
toolkit~\cite{Agostinelli:2002hh}. We involve the cosmic ray spectra dominated
by protons and $^4$He components. We fit the corresponding plot from
PDG\,\cite{Tanabashi:2018oca} by a power-law function, see
Fig.~\ref{fig1}. 
The fits look as (units are m$^{-2}$s$^{-1}$sr$^{-1}$GeV$^{-1}$)
\begin{gather}
  \frac{d^2\Phi}{dE_{kin}d\Omega} = \frac{2.7\times 10^4}{(E_{kin} + 2.3m_p)^{2.78}}\;\;\;
  \;\;\;\text{for protons}\,, \label{protons}\\
  \frac{d^2\Phi}{dE_{kin}d\Omega} = \frac{1.5\times 10^4}{(E_{kin} + 1.4m_{He})^{2.73}}\;\;\;
  \;\;\;\text{for $^4$He}\,,
\end{gather}
where $E_{kin}$ is kinetic energy of the nuclei, $m_p$ and $m_{He}$ are masses of proton and
$^4$He nuclei respectively. In simulations we set 
the threshold kinetic energy $E_{kin}^{th}$ to $0.4$\,GeV for
protons and to 0.8\,GeV for $^4$He. We checked that cosmic rays of lower
energies produce rather small amount of neutrinos.  

The Moon chemical composition is taken as shown in Table~\ref{moon_composition}
\begin{table}[!htb]
  \begin{center}
  \begin{tabular}{|c|c|}
    \hline
    Element & Composition (wt\%) \\
    \hline
    SiO$_2$ & 45.5 \\
    Al$_2$O$_3$ & 19.5\\
    CaO & 13.8 \\
    FeO & 10.0 \\
    MgO & 8.3\\
    TiO$_2$ & 2.3\\
    Na$_2$O & 0.6\\
    \hline
  \end{tabular}
  \caption{\label{moon_composition}Chemical composition
    of lunar soil, adopted from Ref.\,\cite{Moon}.}
  \end{center}
\end{table}
which is an average of chemical compositions of {\it Maria} and {\it
  Highlands} parts of the Moon surface, see e.g.~\cite{Moon}.
For the lunar soil density we take $\rho\approx  1.5$~g$/$cm$^3$
which corresponds to an average density of the upper layer of 
regolith. We assume the isotropic
arrival directions of the cosmic rays. We track the secondary particles produced
by the collisions of cosmic rays with the lunar surface using GEANT (we exploit FTFP\_BERT
physics model). Within the simulation we count all electron and muon neutrinos and
antineutrinos leaving the Moon. Neutrino energy allows for
constructing spectra. The fluxes of $\tau$-neutrinos and antineutrinos
at production 
are negligible as compared to those emerged due to oscillations,
which we discuss below. 

\begin{figure}[!htb]
  \begin{center}
    \begin{tabular}{cc}
      \includegraphics[width=9.0cm]{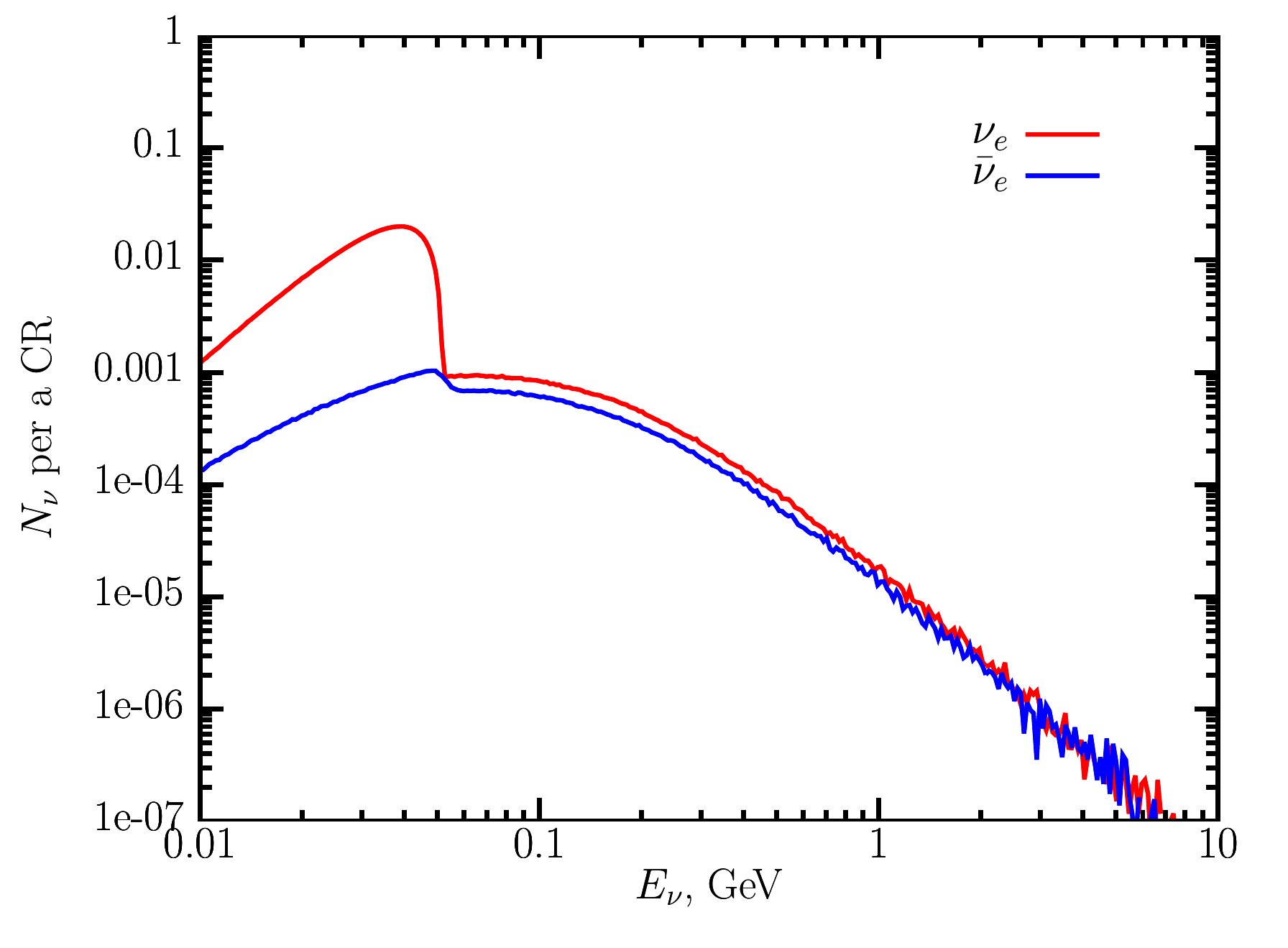} &
      \includegraphics[width=9.0cm]{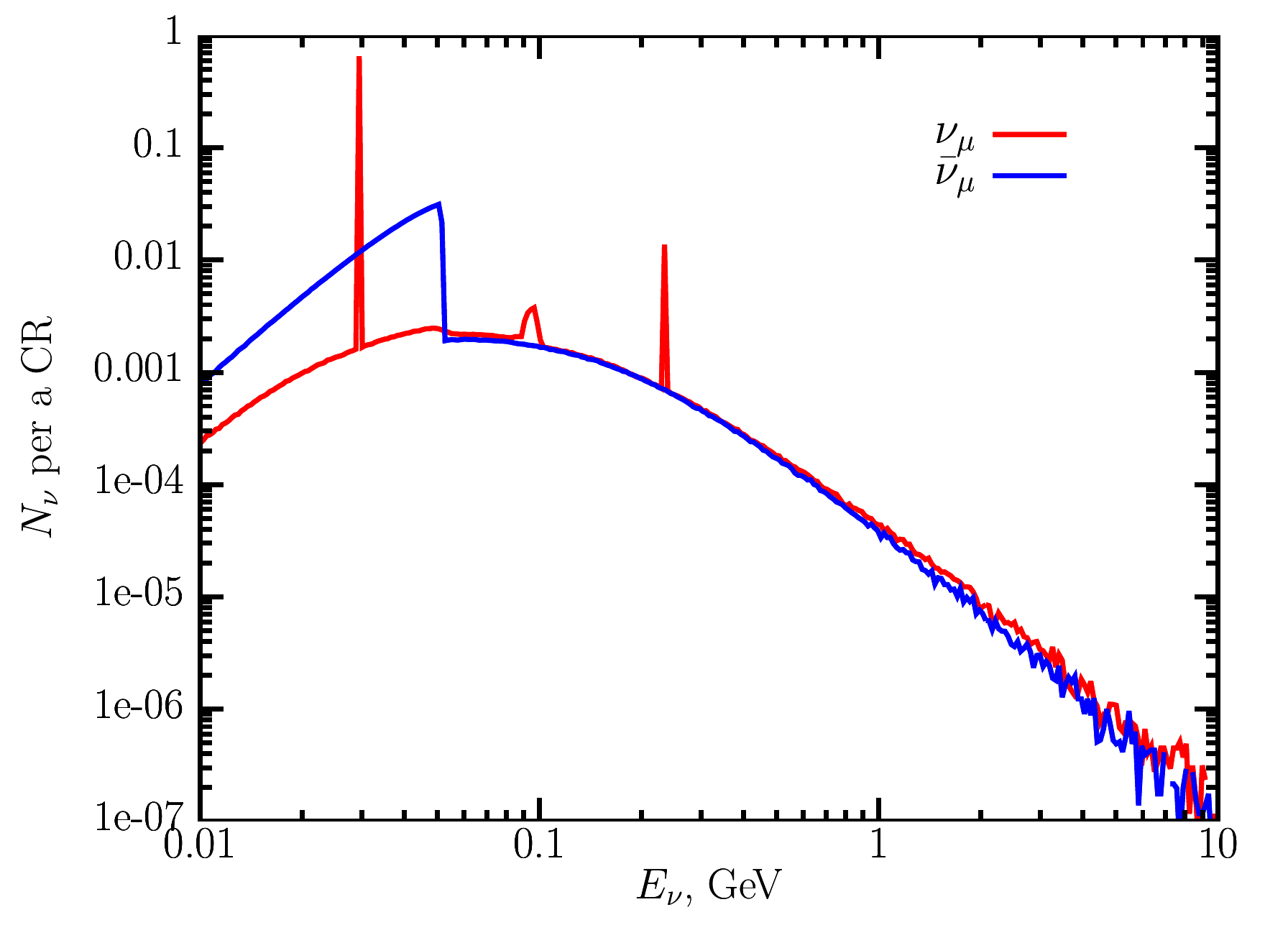}
    \end{tabular}
    \caption{\label{spectrum_nu_per_CR} Spectra of electron (left panel)
      and muon (right panel) neutrinos and antineutrinos normalized to a
      single CR. The energy goes from 10~MeV to 10~GeV in decimal
      logarithmic scale with 320 bins.}
  \end{center}
\end{figure}

In Fig.~\ref{spectrum_nu_per_CR} we show the obtained spectra of
electron (left panel) and muon (right panel) neutrinos  
and antineutrinos $N_\nu(E_\nu)$ normalized to a single cosmic ray hitting the
Moon; we simulated about $1.5\times 10^7$ CR events and resulting neutrino 
events are distributed over 320 uniform logarithmic energy bins.
No oscillations are taken
into account at this stage. Most part of the neutrinos come from charged pion and kaon 
decays as well as from decays of secondary muons. Initially, pions
$\pi^+$ and $\pi^-$ are produced in close 
fractions. They loose their energies in the course of elastic collisions with
nuclei, interact and stop in the lunar soil. Most part of $\pi^-$ get captured
by nuclei via Coulomb attraction~\cite{Ponomarev:1973ya} and do not produce
neutrinos within discussed energy interval 10~MeV -- 10~GeV. On the
other hand, stopped $\pi^+$ produce
a monochromatic line of $\nu_\mu$ at energy $E_\nu\approx 29.8$~MeV
and anitmuons $\mu^+$ which also stop in media and undergo the decay
$\mu^+\to \e^+\nu_e\bar{\nu}_\mu$ yielding neutrinos of 
energies below the threshold at $E_\nu\approx 52.8$~MeV. Similar picture is valid for charged
kaons, which produce a monochromatic $\nu_\mu$ line at $E_\nu\approx
235.6$~MeV. On the right panel of Fig.~\ref{spectrum_nu_per_CR} two peaks at $E_\nu\sim
30$~MeV and~236~MeV in the red histogram correspond to those monochromatic
neutrinos. The widths of 30~MeV
and~236~MeV lines on this plot are chosen to be equal to the bin
width. Their actual widths are considerably narrower, and hence the
actual heights are considerably higher. We clarify their contribution numerically
in due course. 
There is also a small bump at energies somewhat below the muon mass,  
$E_\nu\approx 100$~MeV. It appears from the processes of muon capture
by nuclei with subsequent conversion to neutrino, i.e. 
$\mu^- +
p\to n + \nu_\mu$, see e.g.~\cite{Measday:2001yr}. Spectra of
$\bar{\nu}_\mu$ and $\nu_e$ at $E_\nu\lsim 53$~MeV correspond to neutrinos
from decays of antimuons $\mu^+$, most of which decay at rest. 
Note in passing that neutrinos from stopped pions and kaons were studied
previously in different contexts (and in particular in searches for a signal
from dark matter annihilations in the Sun) in
Refs.~\cite{Spitz:2014hwa,Rott:2012qb,Bernal:2012qh,Rott:2015nma,Rott:2016mzs}. 

Now we turn to calculation of lunar neutrino flux at the Earth.
The lunar regolith emits neutrinos in all directions\footnote{Neglecting small
  cosmic ray shadow by the Earth at the Moon.}. 
Introducing total integrated over energy isotropic flux $\Phi_{CR}$ of cosmic rays 
used in the simulations we can find the number of cosmic ray particles 
bombarding the Moon per second  
\be
\int 2\pi d(\cos{\theta}) \cos{\theta}\, dS \Phi_{CR} =
\pi\cdot 4\pi R_{\rm Moon}^2\Phi_{CR}\,, 
\ee
where $R_{\rm Moon}$ is the Moon radius. Then, a neutrino $\nu_\alpha$
flux at the Earth can be written as
\be
\Phi_{\nu_{\alpha}} = \frac{1}{4\pi L_{\rm Moon}^2} \times \pi\cdot 4\pi R_{\rm
  Moon}^2\Phi_{CR}N_{\nu_{\alpha}}(E_\nu) \equiv \pi \left(\frac{R_{\rm
    Moon}}{L_{\rm Moon}}\right)^2\,\Phi_{CR}\,N_{\nu_{\alpha}}(E_\nu)\,,
\ee
where $L_{\rm Moon}$ is an average distance from the Earth to the Moon.
Let us note that given Moon's orbit perigee about 362600\,km and apogee
about 405400\,km one expects monthly variations of the lunar neutrino
flux $\Phi_{\nu_{\alpha}}$ with the amplitude about 12\%.

\begin{figure}[!htb]
  \begin{center}
    \begin{tabular}{cc}
      \includegraphics[width=9.0cm]{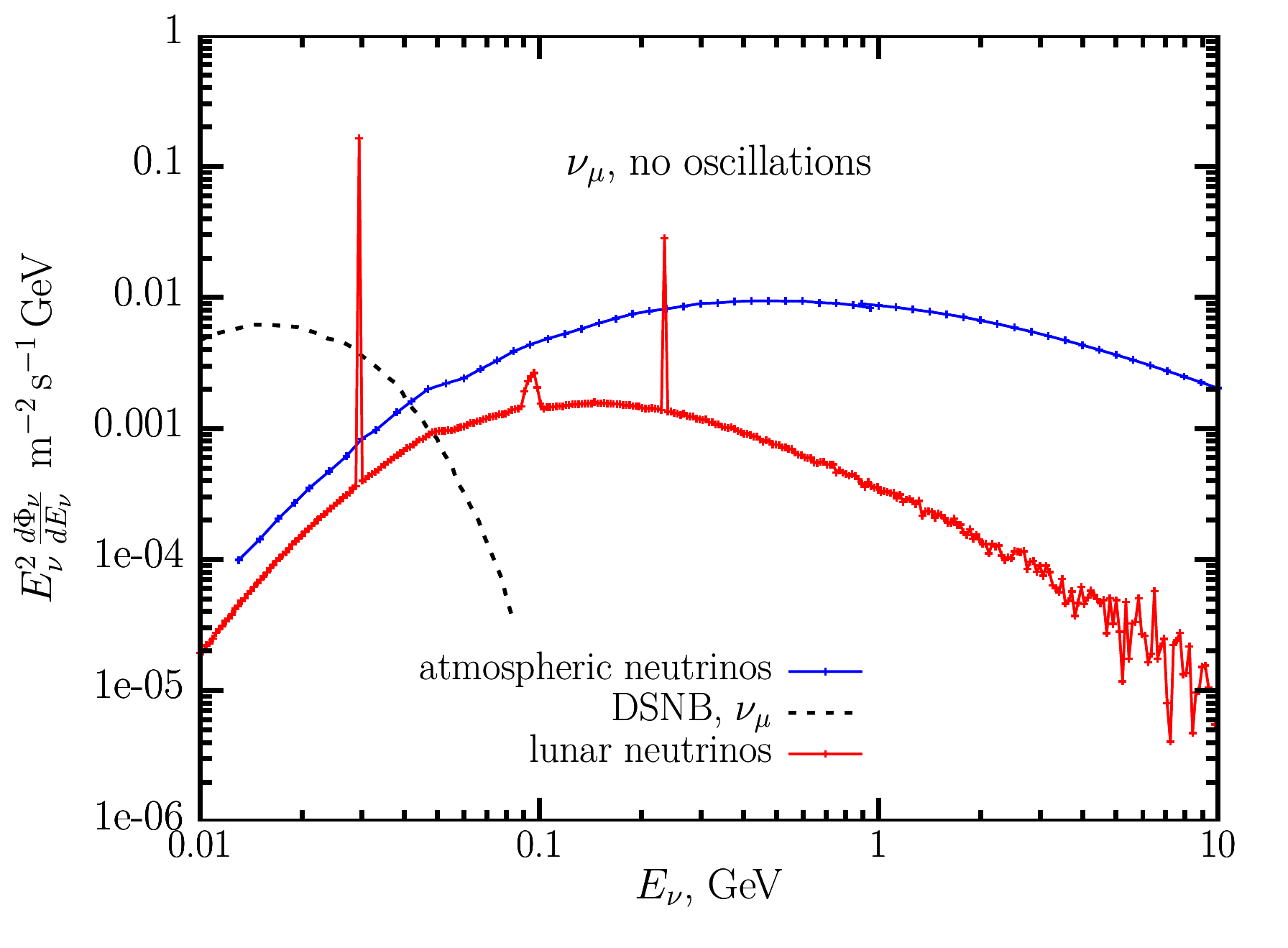} &
      \includegraphics[width=9.0cm]{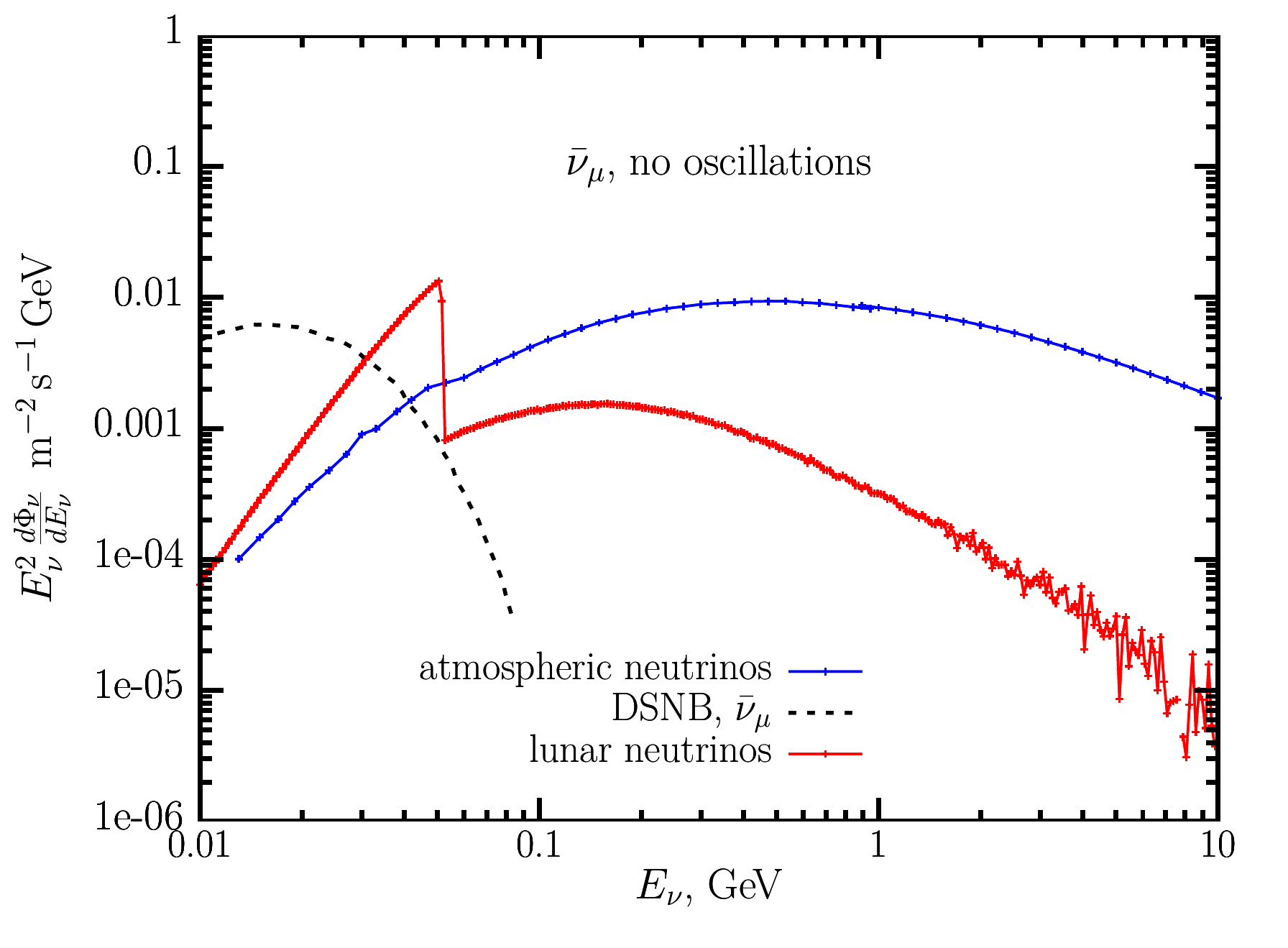}
    \end{tabular}
    \caption{\label{dFlux_numu_anumu} Fluxes of lunar neutrinos $\nu_\mu$
      and antineutrinos $\bar{\nu}_\mu$ (red lines) in comparison with
      atmospheric (blue) and diffuse supernova (dashed black) neutrino fluxes
      calculated within a fraction of solid 
      angle $\frac{\pi R_{\rm Moon}^2}{4\pi L_{\rm Moon}^2}$.
      No oscillations are taken into account.}
  \end{center}
\end{figure}

\begin{figure}[!htb]
  \begin{center}
    \begin{tabular}{cc}
      \includegraphics[width=9.0cm]{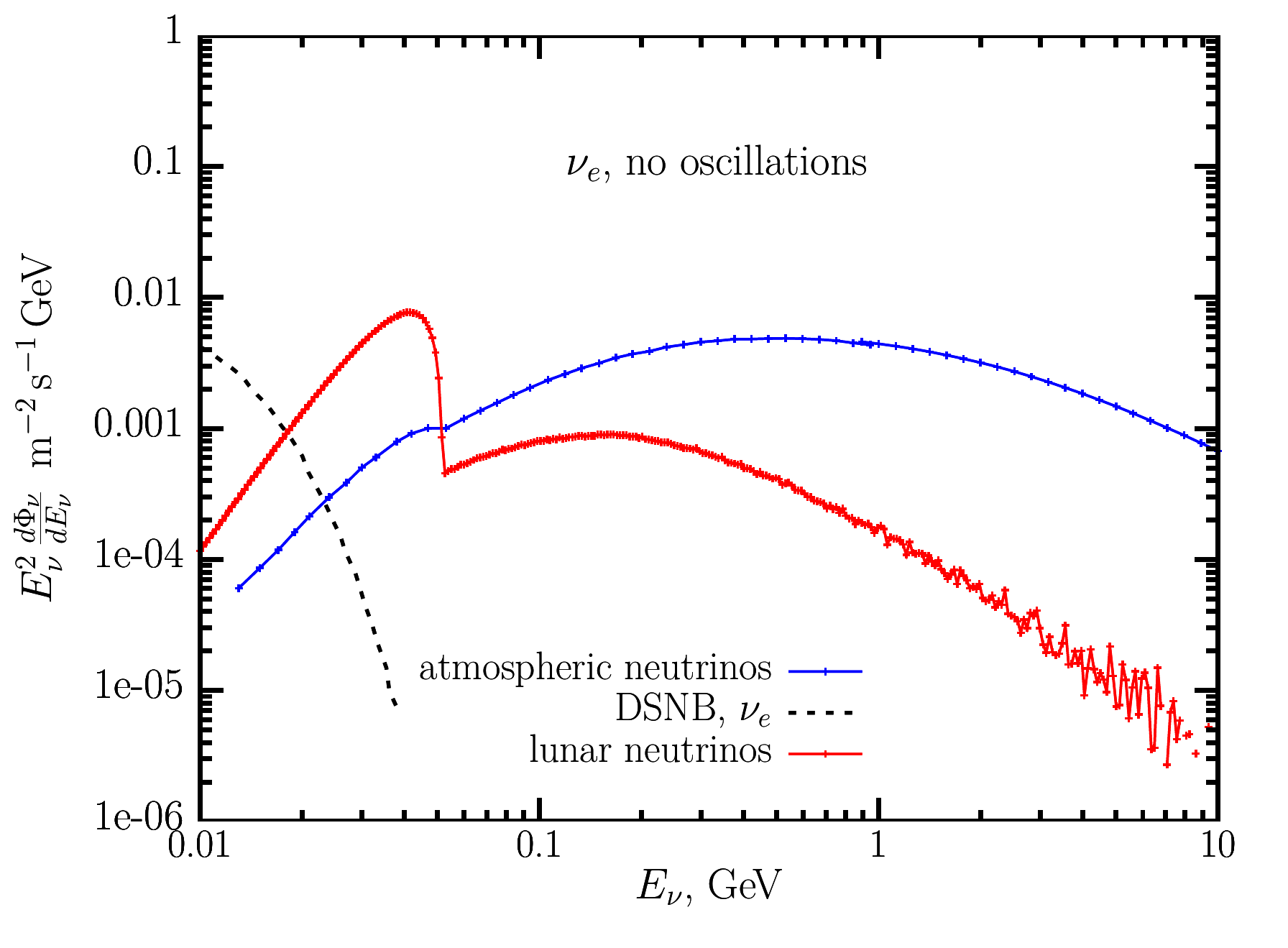} &
      \includegraphics[width=9.0cm]{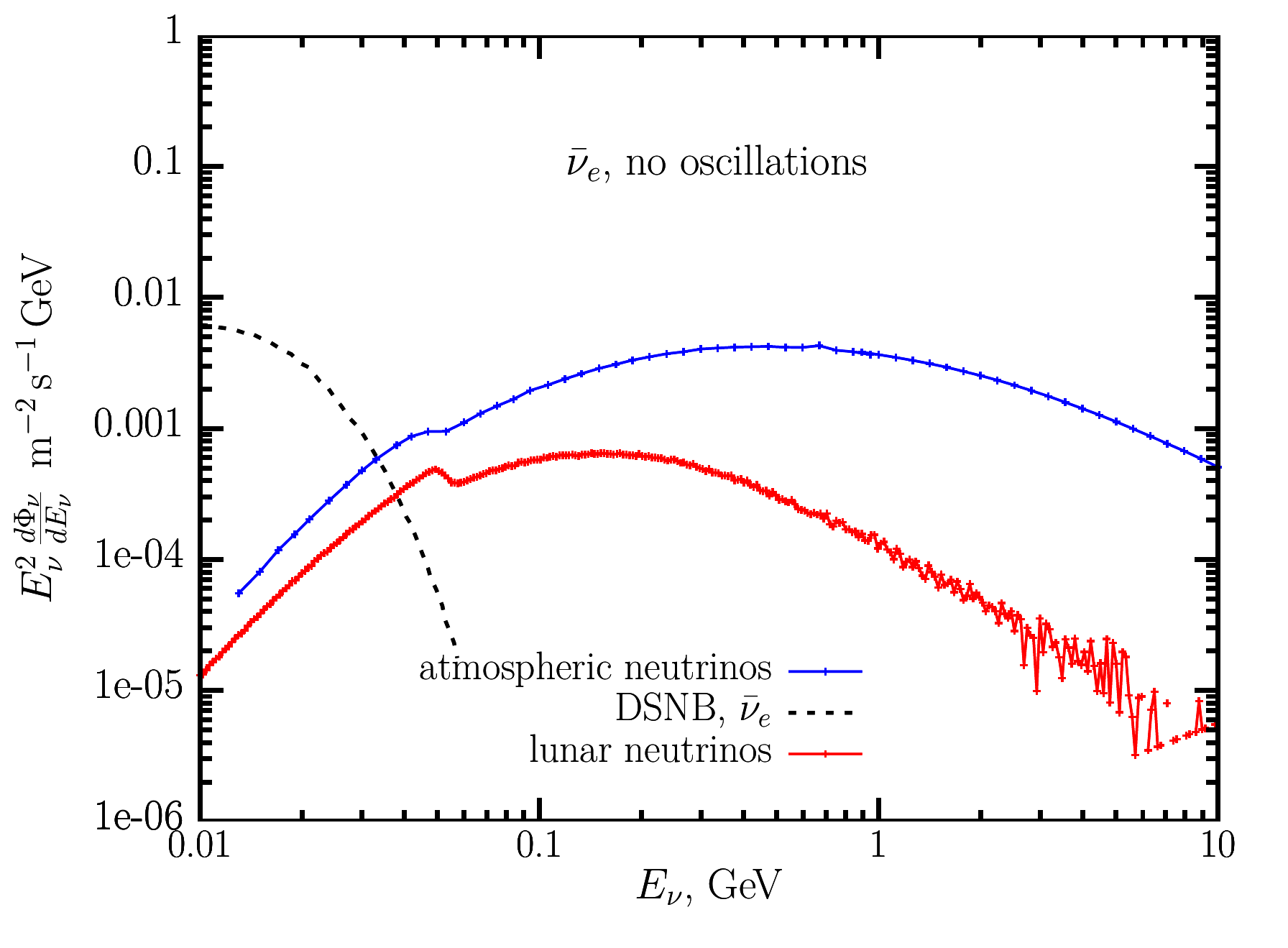}
    \end{tabular}
    \caption{\label{dFlux_nue_anue} Fluxes of lunar neutrinos $\nu_e$
      and anitneutrinos $\bar{\nu}_e$ (red lines) in comparison with
      atmospheric (blue) and diffuse supernova (dashed black) neutrino fluxes
      calculated within a fraction of solid
      angle $\frac{\pi R_{\rm Moon}^2}{4\pi L_{\rm Moon}^2}$.
      No oscillations are taken into account.}
  \end{center}
\end{figure}

In Figs.~\ref{dFlux_numu_anumu} and~\ref{dFlux_nue_anue} 
we present differential neutrino fluxes multiplied by energy squared,
i.e. $E_\nu^2\frac{d\Phi_{\nu}}{dE_\nu}$
(in units of m$^{-2}\,$s$^{-1}\,$GeV), for muon and electron
(anti)neutrinos (red lines) in comparison with atmospheric
neutrino 
flux (blue lines) taken from\,\cite{Battistoni:2005pd,Honda:2015fha} and
  diffuse supernova neutrino background (DSNB) flux (dashed black lines)
taken from~\cite{OHare:2015utx}  in the direction towards the Moon,
which are obtained\footnote{Let us note that
that atmospheric neutrino flux depends on the positioning of the
detector, the difference is 
within a factor of two~\cite{Battistoni:2005pd}.}
with multiplying the overall neutrino fluxes (no oscillations)
by a factor equal to the fraction of the celestial sphere occupied by
the Moon, $\frac{\Delta\Omega}{4\pi} = \frac{\pi R_{\rm
    Moon}^2}{4\pi L_{\rm Moon}^2}\approx 5\times 10^{-6}$. 
Comparing, for instance, lunar and atmospheric neutrinos in the energy
interval from 10~MeV to 1~GeV we find that ratio of their energy
integrated fluxes within the
same solid angle is close to unity (about 1.4--1.5). At the same time,
shapes 
of their energy spectra are drastically different due to prominent features
related to peculiarities of neutrino production in the Moon. In
particular, for an idealised neutrino detector having energy
resolution of 10\% the fluxes of $\nu_\mu$ from the Moon are about
0.14~and 0.0033~m$^{-2}$s$^{-1}$ for neutrino energies $29.8$ and
236~MeV, respectively, to be compared with atmospheric $\nu_\mu$
neutrino fluxes 0.0028 and 0.0034~m$^{-2}$s$^{-1}$, calculated over
the solid angle of the Moon (without oscillations). 

\begin{figure}[!htb]
  \begin{center}
    \begin{tabular}{cc}
      \includegraphics[width=9.0cm]{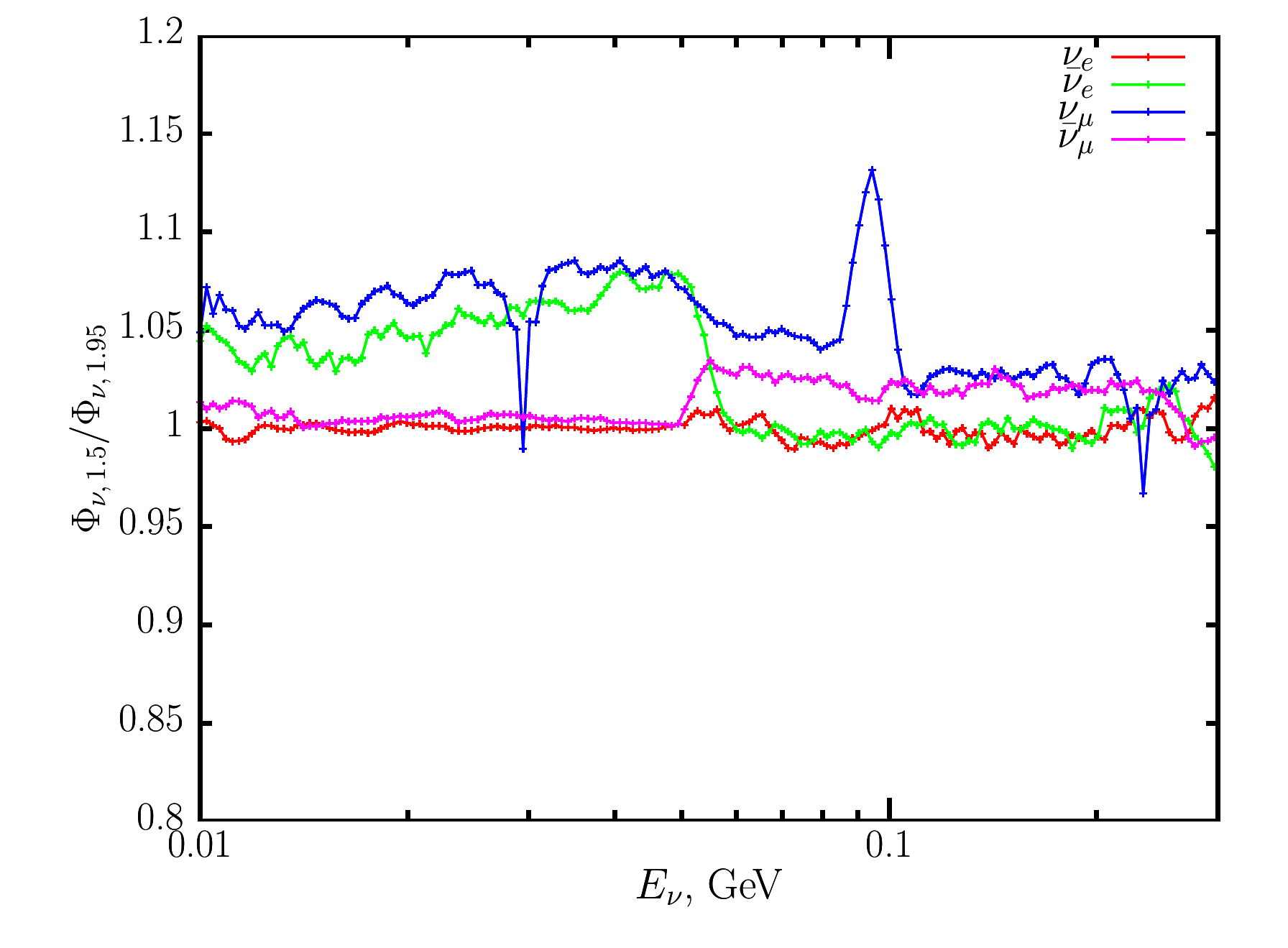} &
    \end{tabular}
    \caption{\label{dFlux_ratio} Ratio of lunar neutrino flux at
      production calculated for different regolith densities, 1.5~and
      1.95~g/cm$^3$, respectively.}   
  \end{center}
\end{figure}

As we discussed above interactions of parent mesons and muons
in the media have a great impact on spectra of 
neutrinos produced in collisions of cosmic rays with the Moon
surface. To illustrate this point we perform the same numerical
simulation taking somewhat larger values of the regolith density,
$\rho=1.95$~g$/$cm$^3$. In Fig.~\ref{dFlux_ratio} 
we show the ratios of neutrino fluxes with densities 1.5 and
1.95~g$/$cm$^3$, respectively, for all neutrino flavors at
production. We see the change of density has the most prominent impact
on $\nu_\mu$ flux. In particular, in the matter of lower density the
flux of $\nu_\mu$ is higher away from peaks at 29.7~MeV and
236~MeV. This corresponds to an increase of probability for mesons 
$\pi^+$ and $K^+$ (and subsequent $\mu^+$) to decay in flight. At
the same time, $\nu_\mu$ flux at the resonant energies, corresponding
to 
decays of mesons at rest, decreases. Also we observe an increase of
$\nu_\mu$ flux at the region around 100~MeV where these neutrinos
appear from $\mu^--\nu_\mu$ conversion on nuclei.
Fluxes of $\bar{\nu}_e$ are also higher at energies corresponding to
neutrino production from stopped muons $\mu^-$. It is
related to lower probability for $\pi^-$ to be captured by nuclei in
less dense media. At the same time fluxes of $\nu_e$ and
$\bar{\nu}_\mu$ which come from decays of $\mu^+$ at this energy
range are almost unchanged. Note that we omit the part of the plot at
larger energies where small statistics do not allow to see clearly
the density impact. It is worth to note that most neutrinos in
the interesting energy range are produced in the surface upper layer
of $\lsim 1$\,m depth. One can also expect dependence of lunar
neutrino flux as on the density as well as on the composition of the
regolith. Hence, the measurement of the neutrino spectra
provides with {\it a tool to investigate the lunar surface density}.

\begin{figure}[!htb]
  \begin{center}
    \begin{tabular}{cc}
      \includegraphics[width=9.0cm]{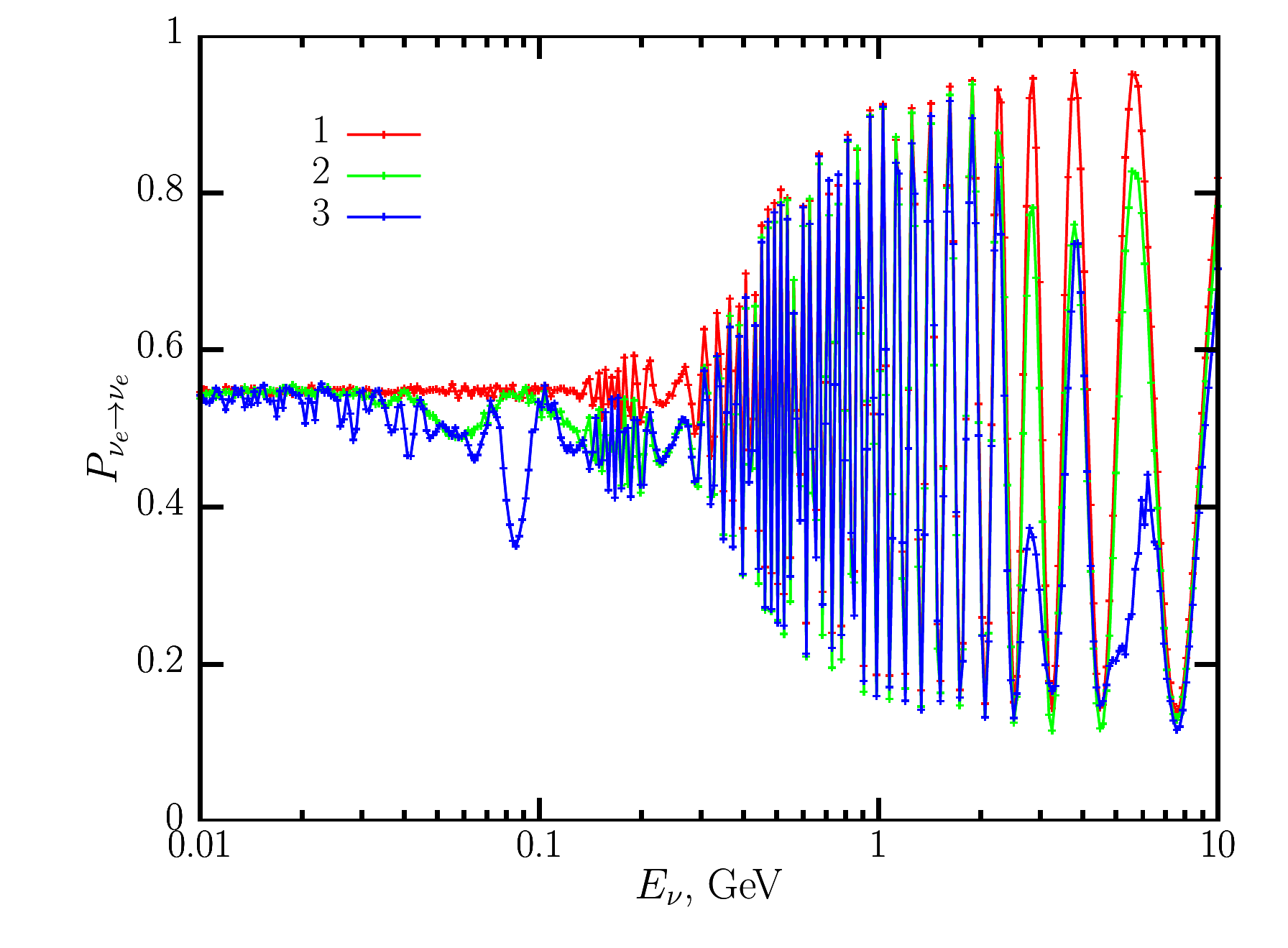} &
      \includegraphics[width=9.0cm]{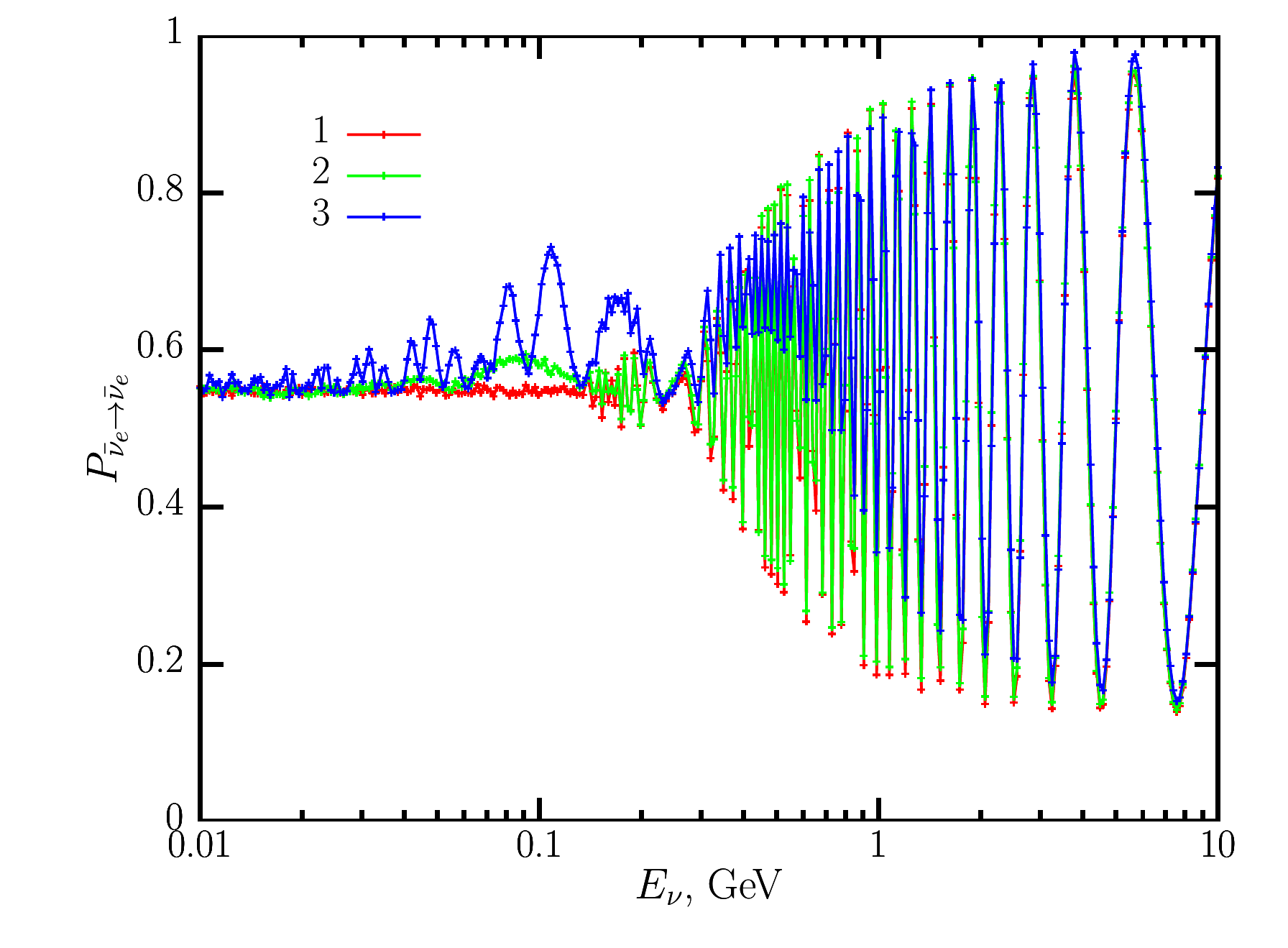} \\
      \includegraphics[width=9.0cm]{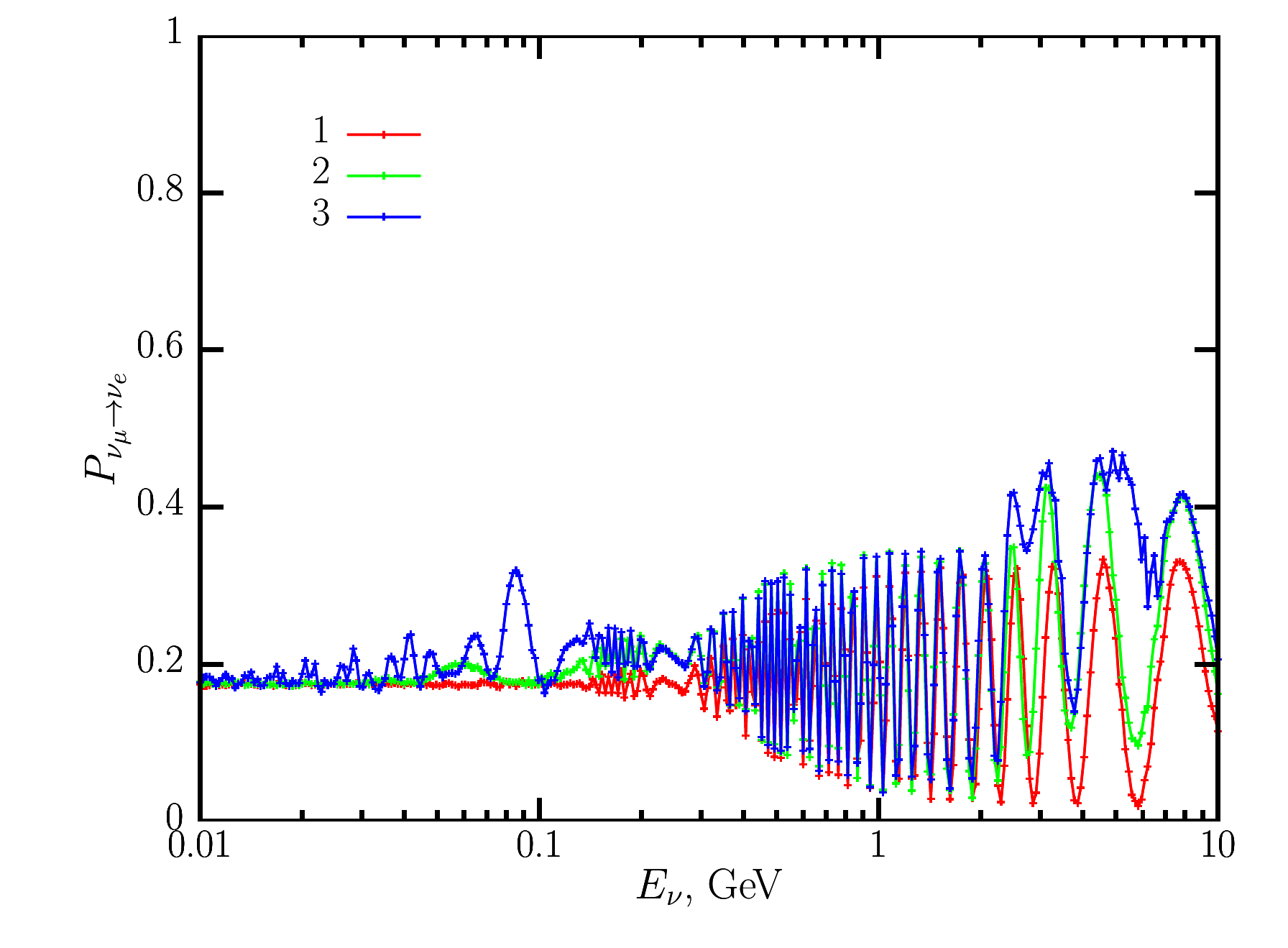} &
      \includegraphics[width=9.0cm]{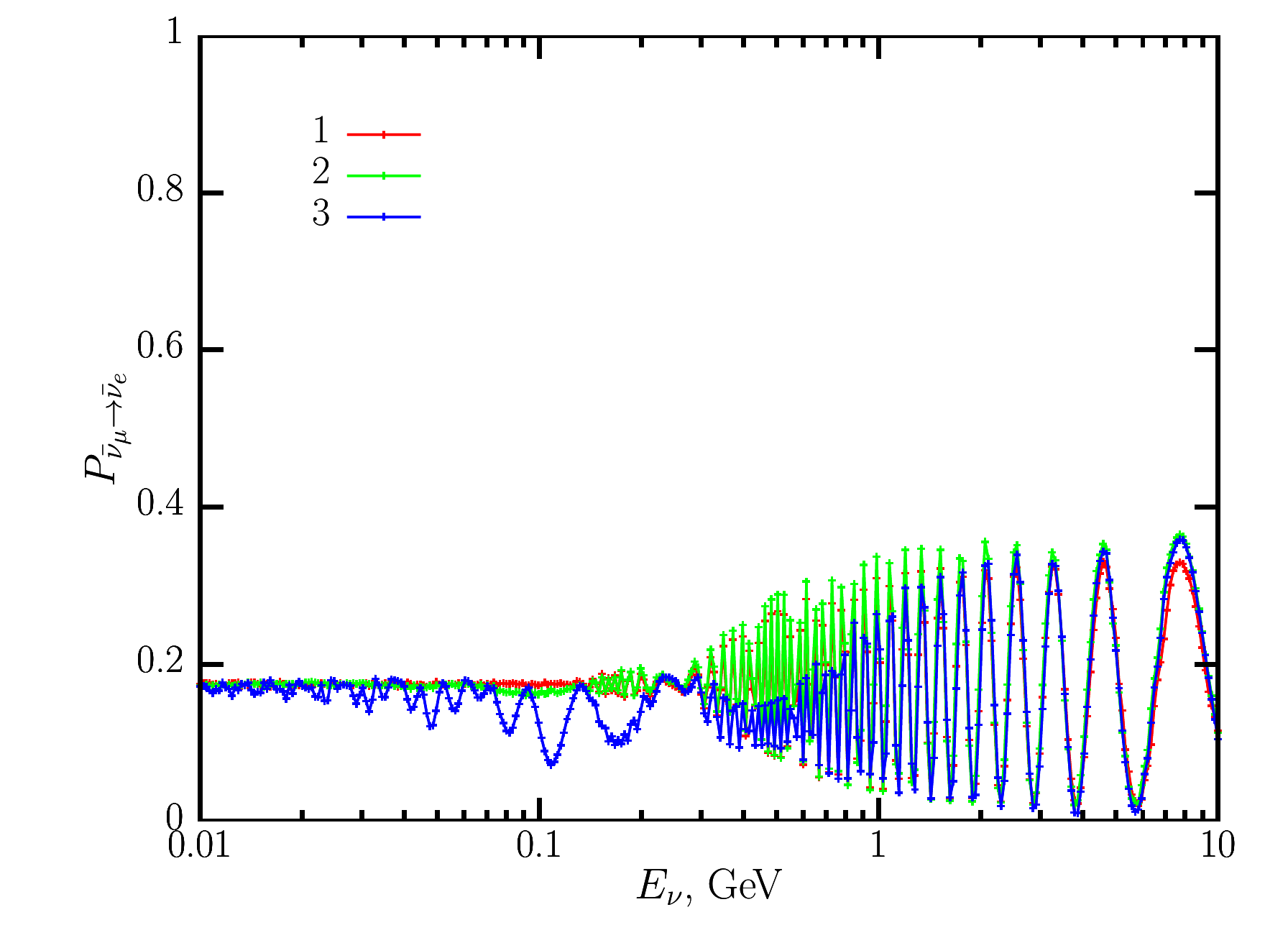}
    \end{tabular}
    \caption{\label{osc_1} Oscillation probabilities of lunar
    neutrinos at the Earth, $P_{\nu_e\to\nu_e}$ (upper left),
    $P_{\bar{\nu}_e\to\bar{\nu}_e}$ (upper right),
    $P_{\nu_\mu\to\nu_e}$ (lower left),
    $P_{\bar{\nu}_\mu\to\bar{\nu}_e}$ (lower right). Different curves
    correspond to probabilities calculated 1) without matter effects;
    2) with matter effects in the Moon only; 3) with matter effects in
    the Moon and the Earth. See main text for details.} 
  \end{center}
\end{figure}

\begin{figure}[!htb]
  \begin{center}
    \begin{tabular}{cc}
      \includegraphics[width=9.0cm]{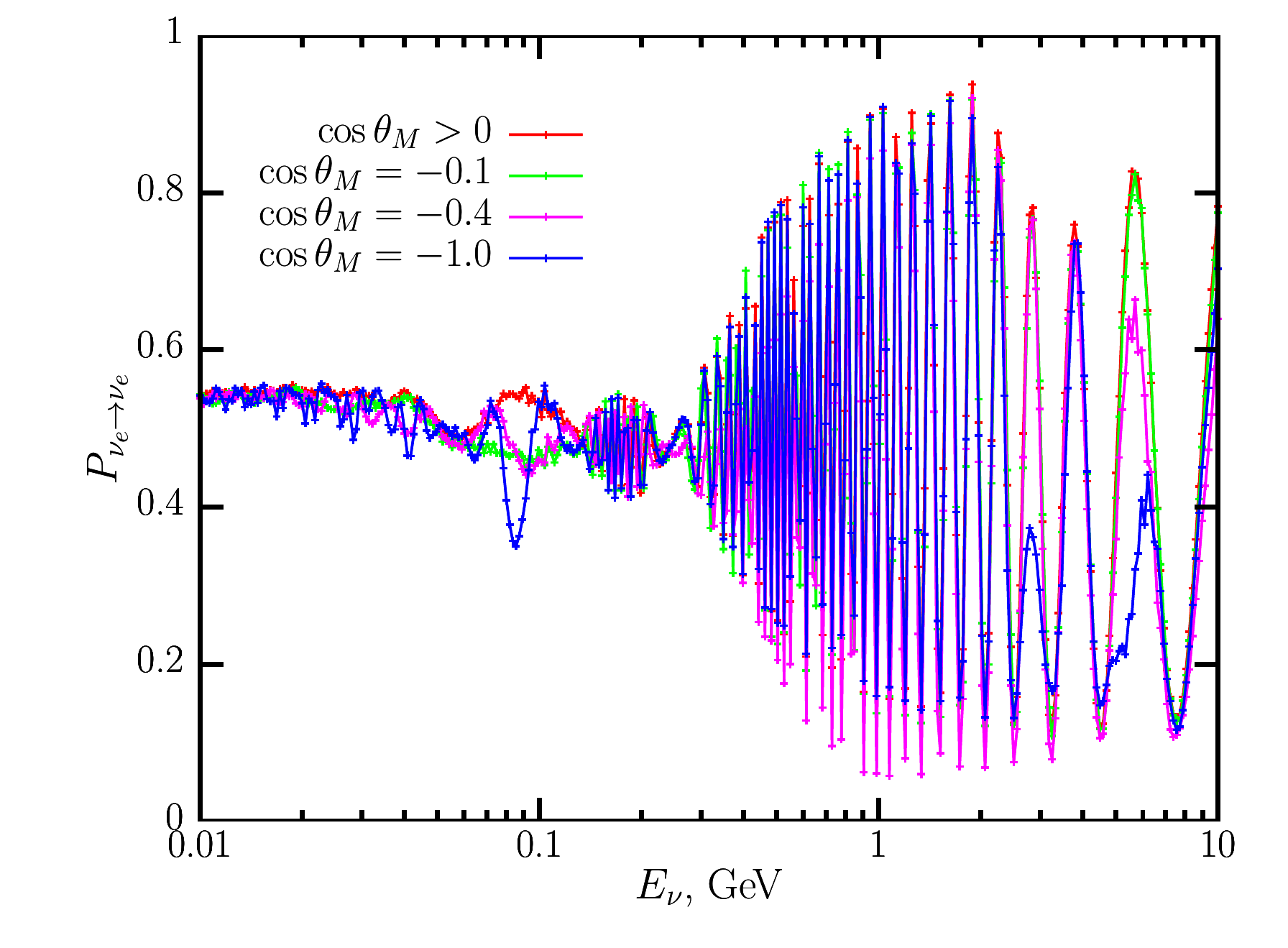} &
      \includegraphics[width=9.0cm]{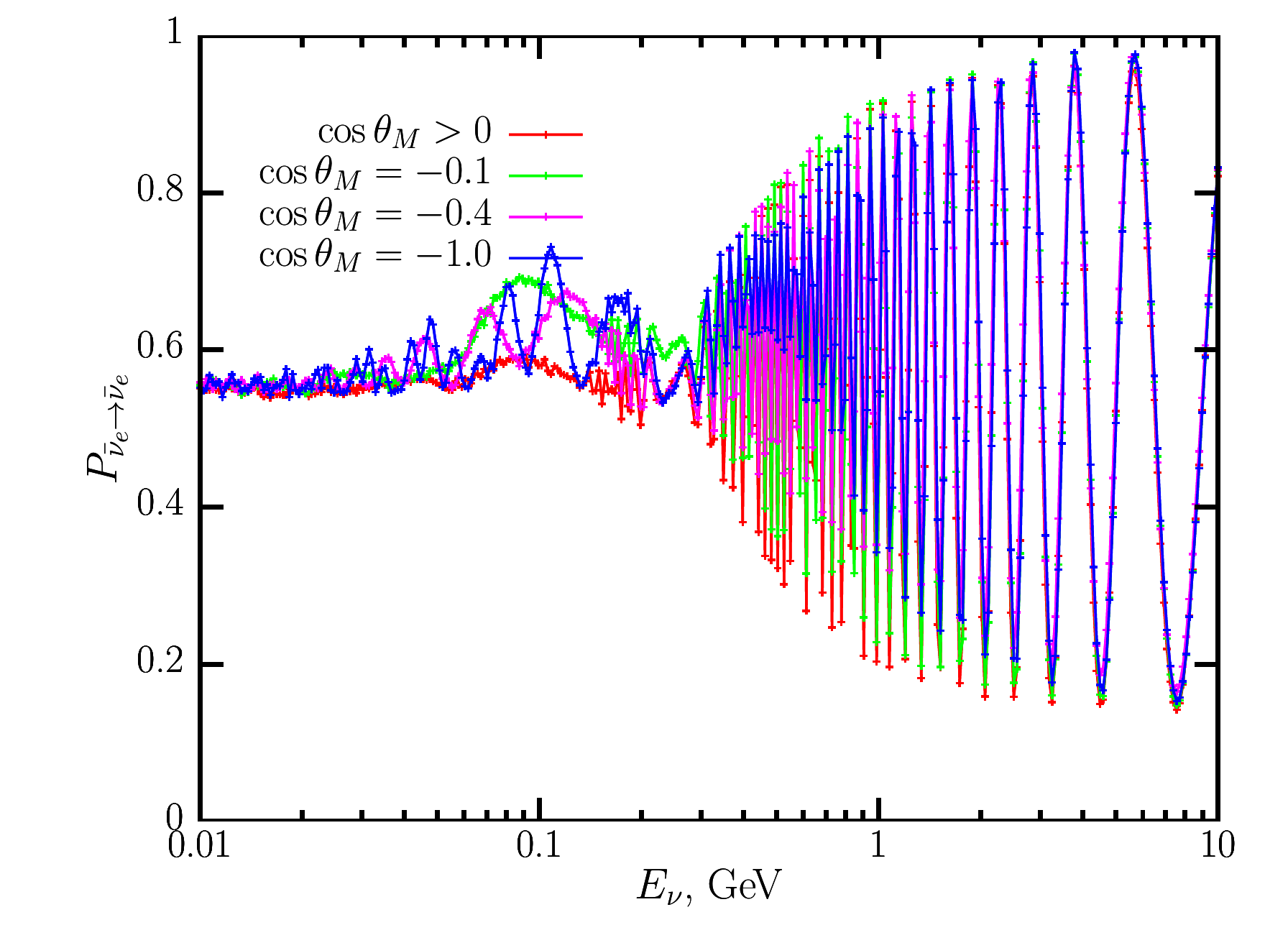} \\
      \includegraphics[width=9.0cm]{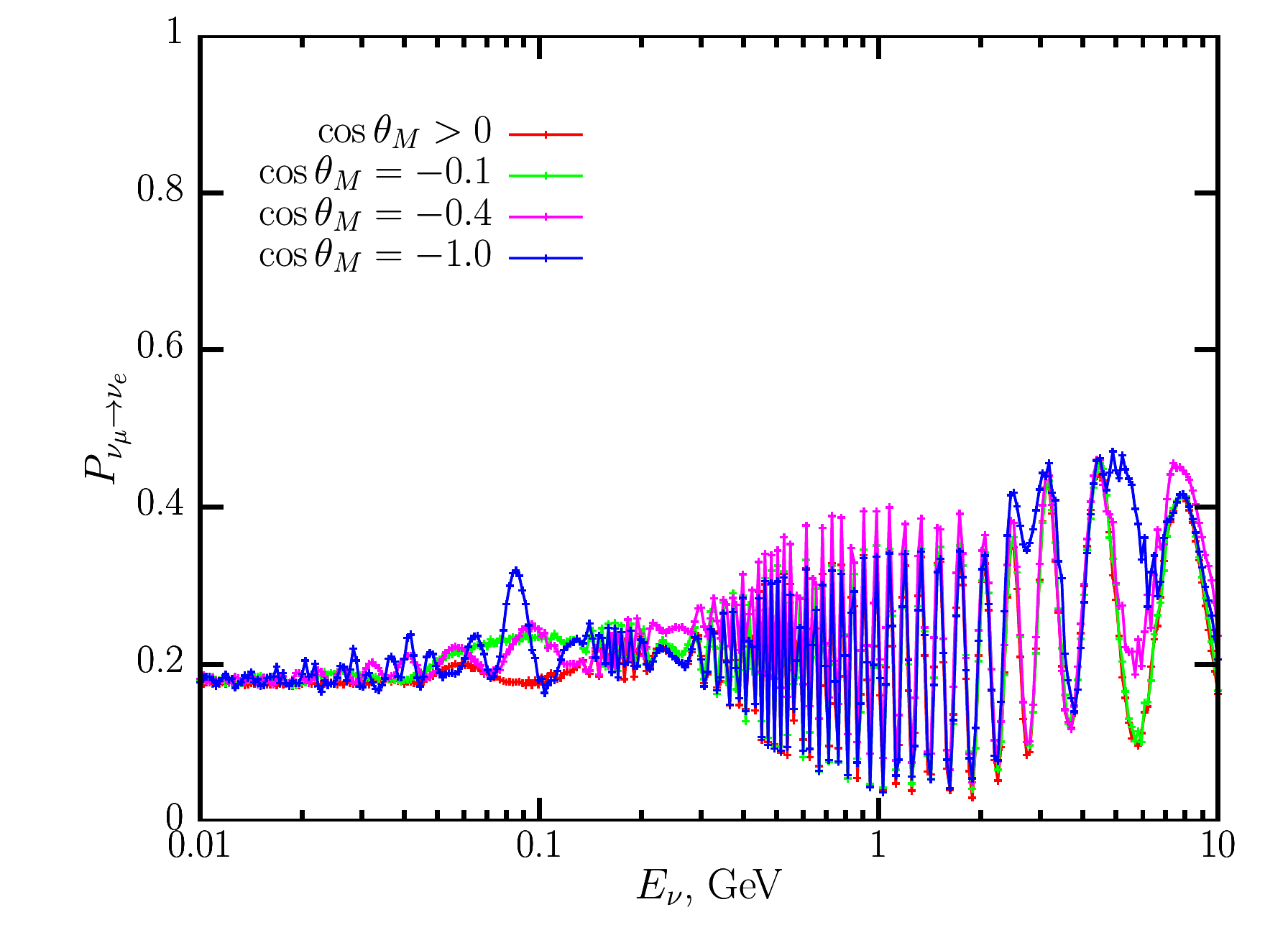} &
      \includegraphics[width=9.0cm]{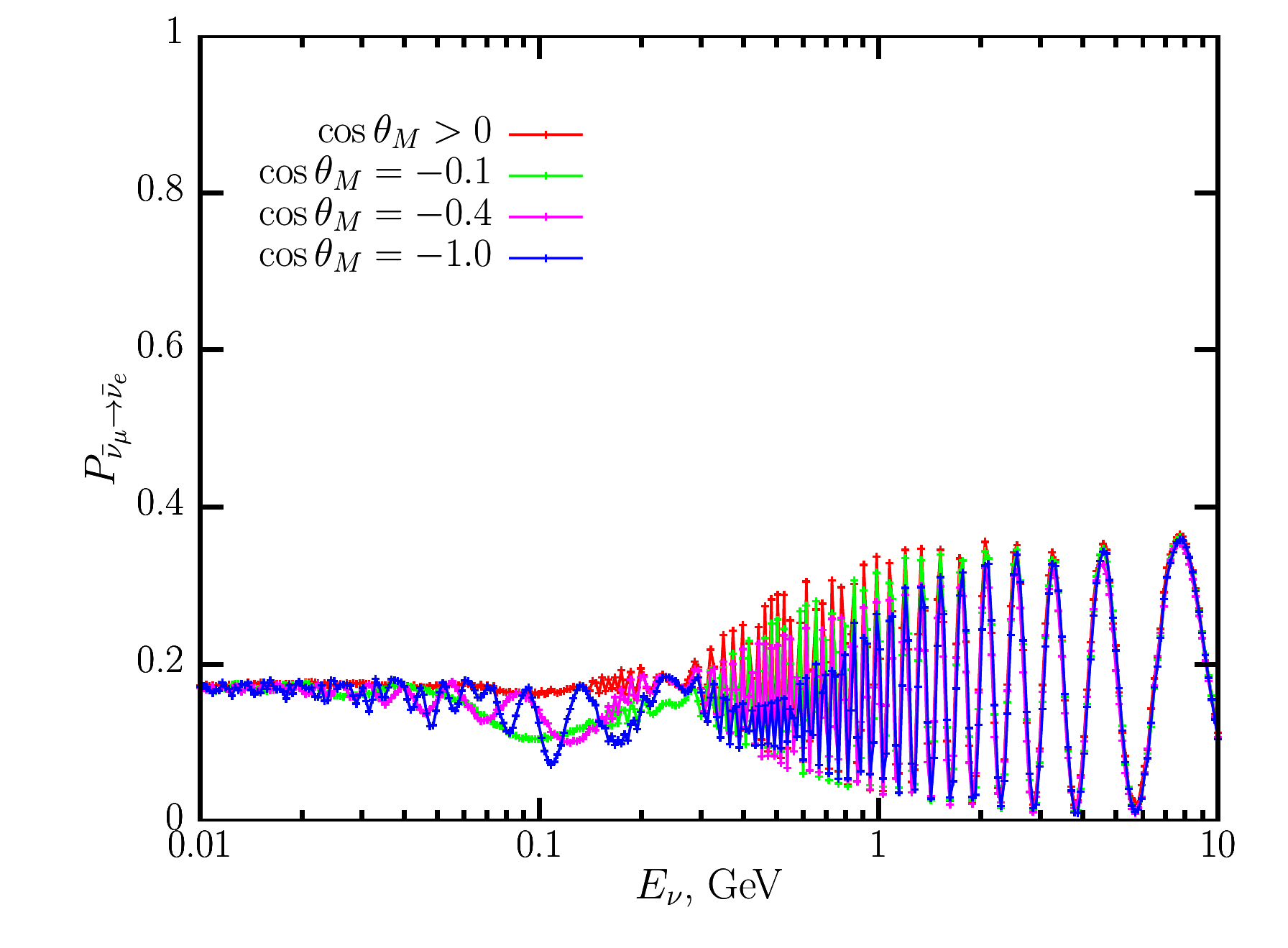}
    \end{tabular}
    \caption{\label{osc_2} Oscillation probabilities to electron
    (anti)neutrinos at the Earth for different zenith angles of the
    the Moon. }
  \end{center}
\end{figure}

\begin{figure}[!htb]
  \begin{center}
    \begin{tabular}{cc}
      \includegraphics[width=9.0cm]{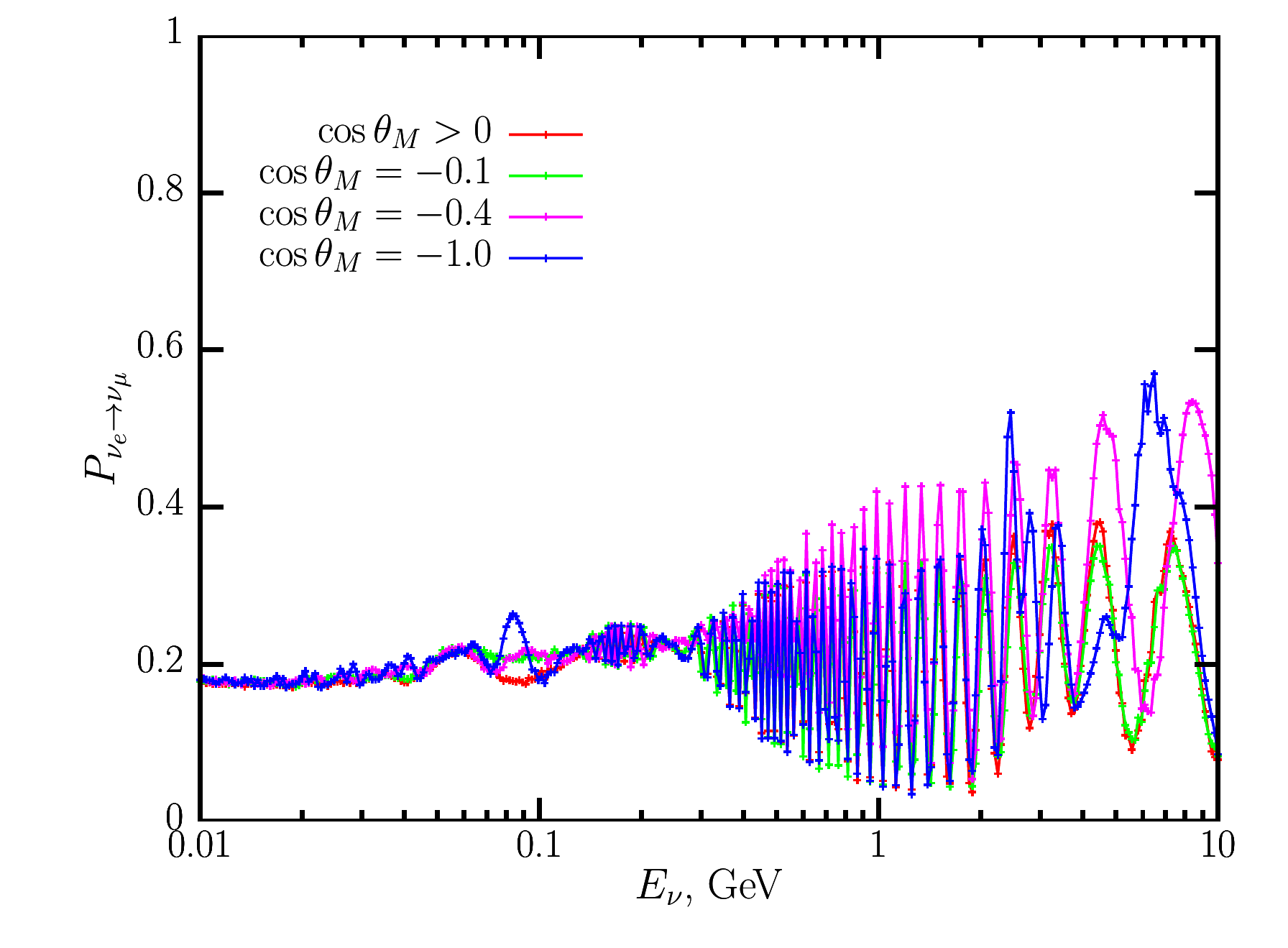} &
      \includegraphics[width=9.0cm]{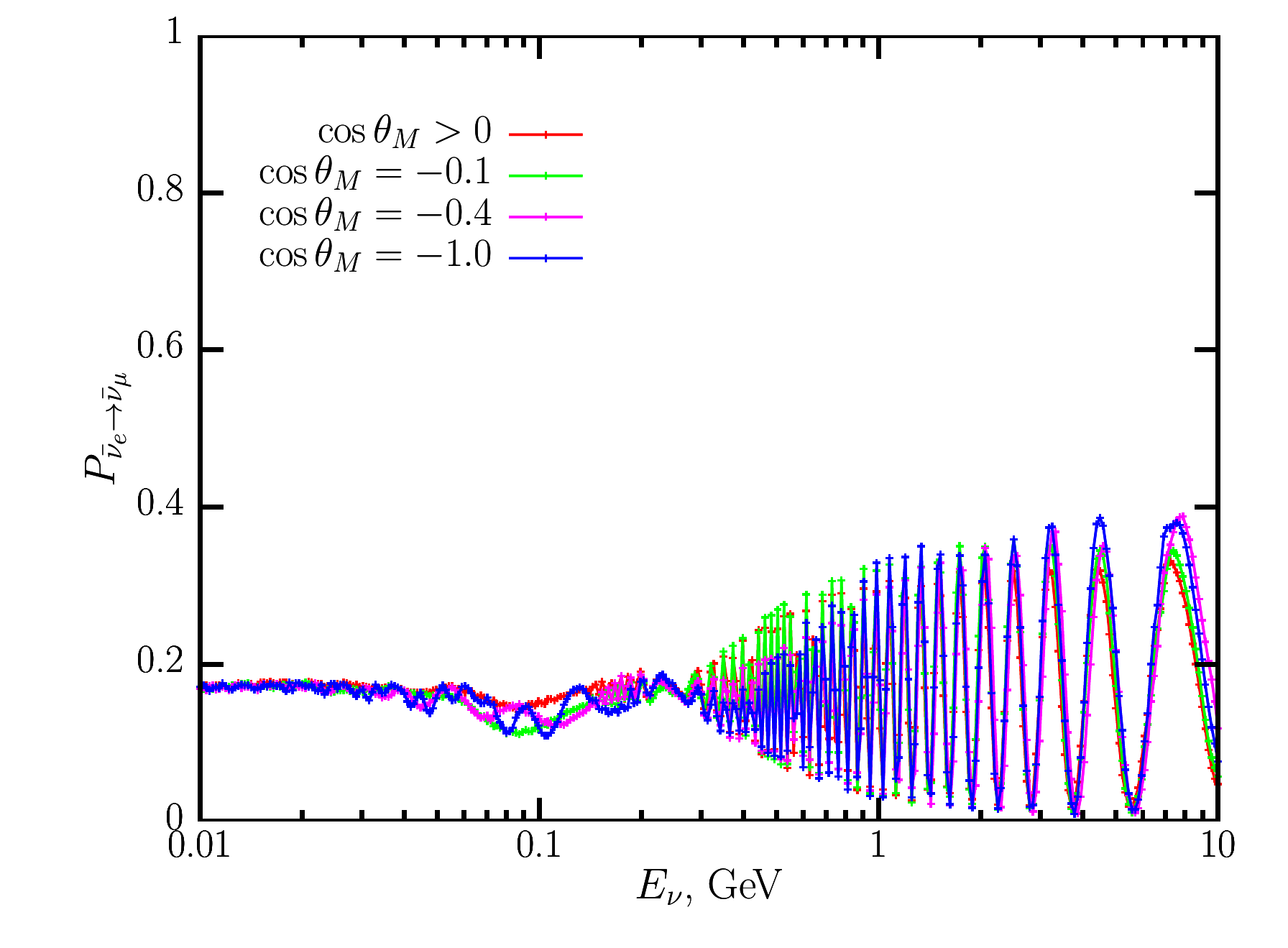} \\
      \includegraphics[width=9.0cm]{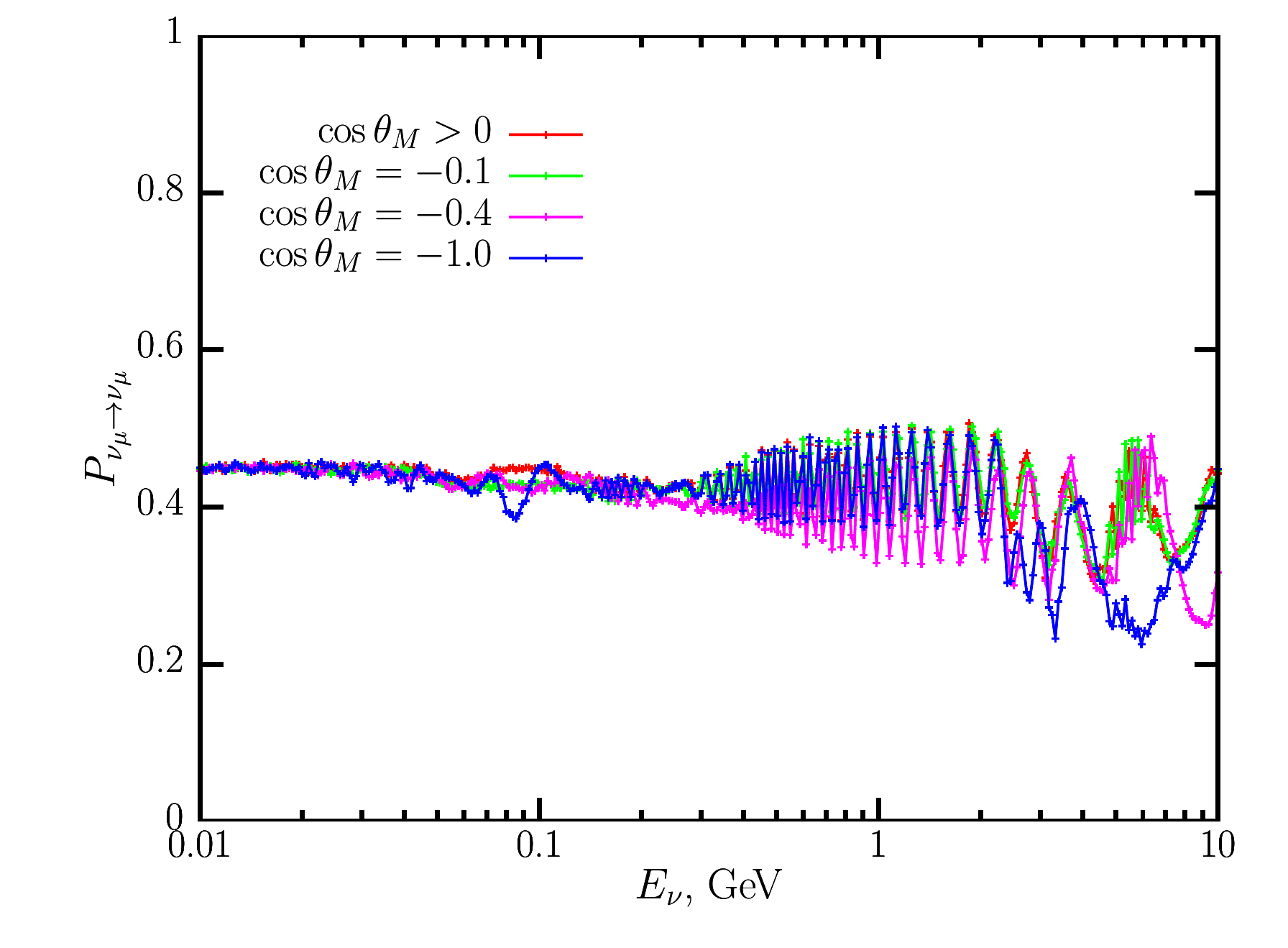} &
      \includegraphics[width=9.0cm]{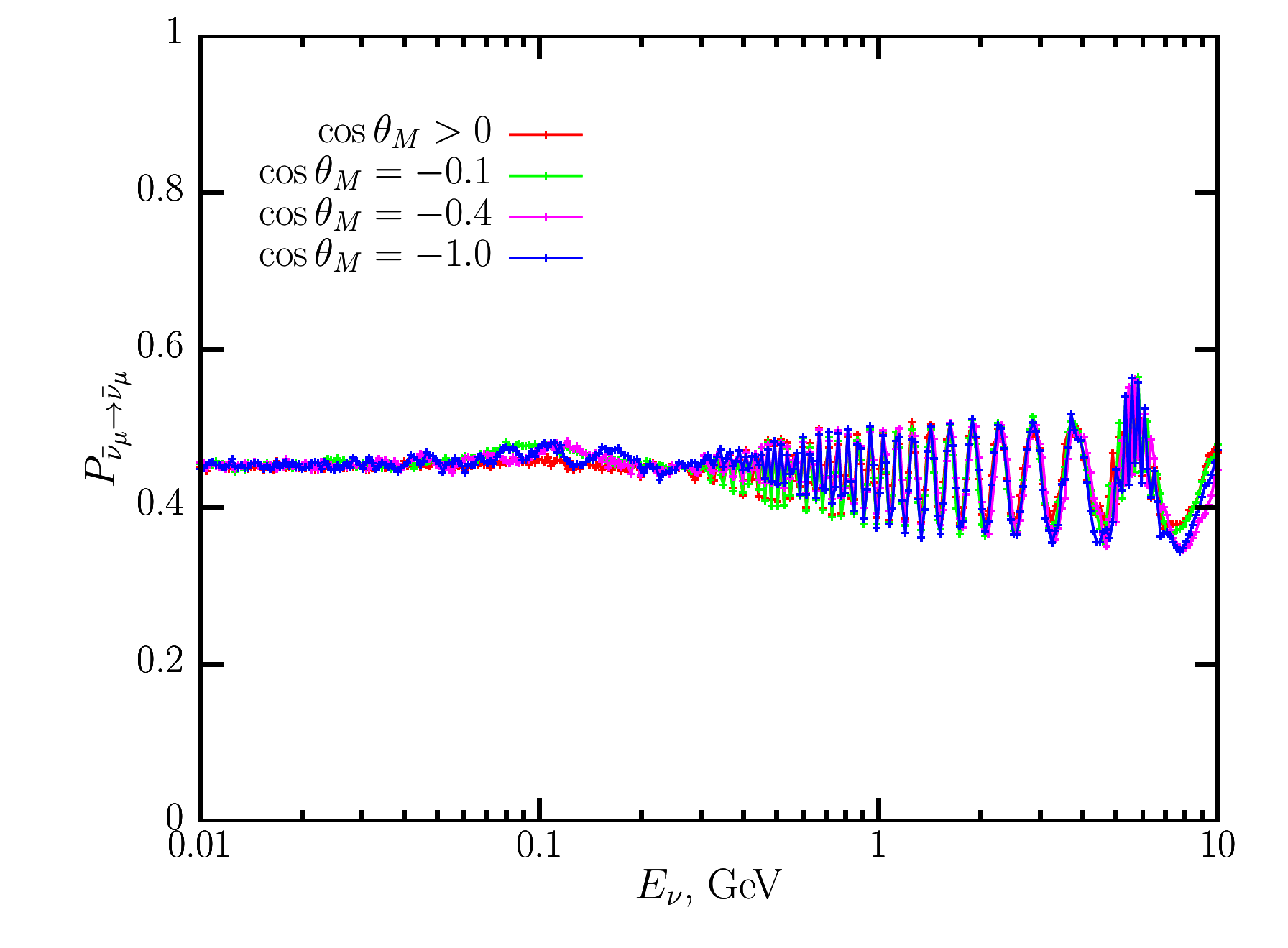}
    \end{tabular}
    \caption{\label{osc_3} Oscillation probabilities for muon
    (anti)neutrinos at the Earth for different zenith angles of the
    Moon.}
  \end{center}
\end{figure}

Now let us study impact of neutrino oscillations on lunar
neutrinos.  As most of them are produced in the surface region of the
Moon, the corresponding neutrino oscillation probabilities should be
averaged over the production point. The oscillation lengths for
neutrinos of 10~MeV --\,1~GeV is considerably smaller than the
baseline which is about $L_{\rm Moon}$ and for lower part of this
energy range is smaller than even radius of the Moon. This results in
almost incoherent neutrino flavour transitions for the softest part of
lunar neutrino spectra. However, matter effects in the Moon and the
Earth may be important\footnote{We are grateful to an anonymous
referee for raising this question.}. To model neutrino oscillation in
the Moon we use profile of lunar density from
Ref.\cite{https://doi.org/10.1002/2013JE004559} and we adopt PREM
model~\cite{DZIEWONSKI1981297} for the Earth structure to describe
neutrino oscillations in the Earth.  We numerically solve the
Schrodinger equation for neutrino wave-function.  For the PMNS matrix
elements we take the best fit to neutrino oscillation experiments from
Ref.~\cite{Esteban:2018azc} for normal neutrino mass hierarchy as an
example. First, to demonstrate the impact of the neutrino oscillation in
matter of the Moon and the Earth we show in Fig.~\ref{osc_1}
neutrino oscillation probabilities $P_{\nu_e\to\nu_e}$ (upper left),
$P_{\bar{\nu}_e\to\bar{\nu}_e}$ (upper right), $P_{\nu_\mu\to\nu_e}$
(lower left), $P_{\bar{\nu}_\mu\to\bar{\nu}_e}$ (lower right) for
three different cases: 1) no matter effects in the Moon and the Earth;
2) with matter effects in the Moon only; 3) matter effects in the Moon
and the Earth. In the latter case we consider neutrino propagating
through the center of the Earth, which implies in a real experiment
that we account for the neutrino signal only when the Moon is in
Nadir. 
In all three cases we average over
the production point on the Moon surface and took $L_{Moon}=384000$~km
as a distance between the Earth and the Moon. We checked that
variation of $L_{Moon}$ due to ellipticity of the Moon orbit results
only in minor corrections to the oscillation probabilities. Also we
average over neutrino energy inside  each of 320 energy bins.
Second, in Fig.~\ref{osc_2}
we show dependence of oscillation probabilities for electron
(anti)neutrinos on zenith angle $\theta_M$ of the Moon (the case
$\cos{\theta_M}>0$ corresponds to the Moon above the horizon, and so
no matter effect in the Earth). In
Fig.~\ref{osc_3} 
we show the same probabilities but to muon (anti)neutrinos. 

\begin{figure}[!htb]
  \begin{center}
    \begin{tabular}{cc}
      \includegraphics[width=9.0cm]{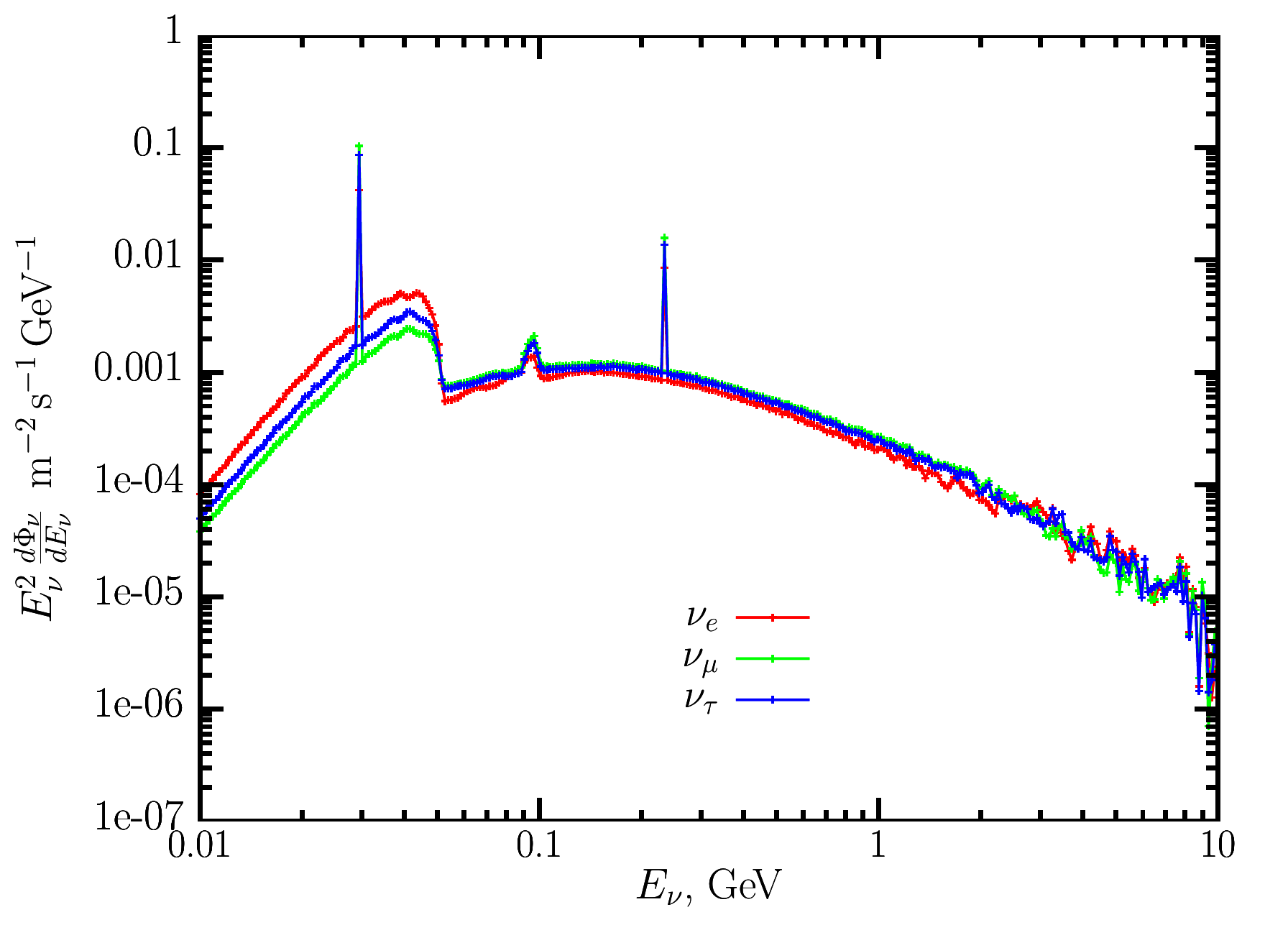} &
      \includegraphics[width=9.0cm]{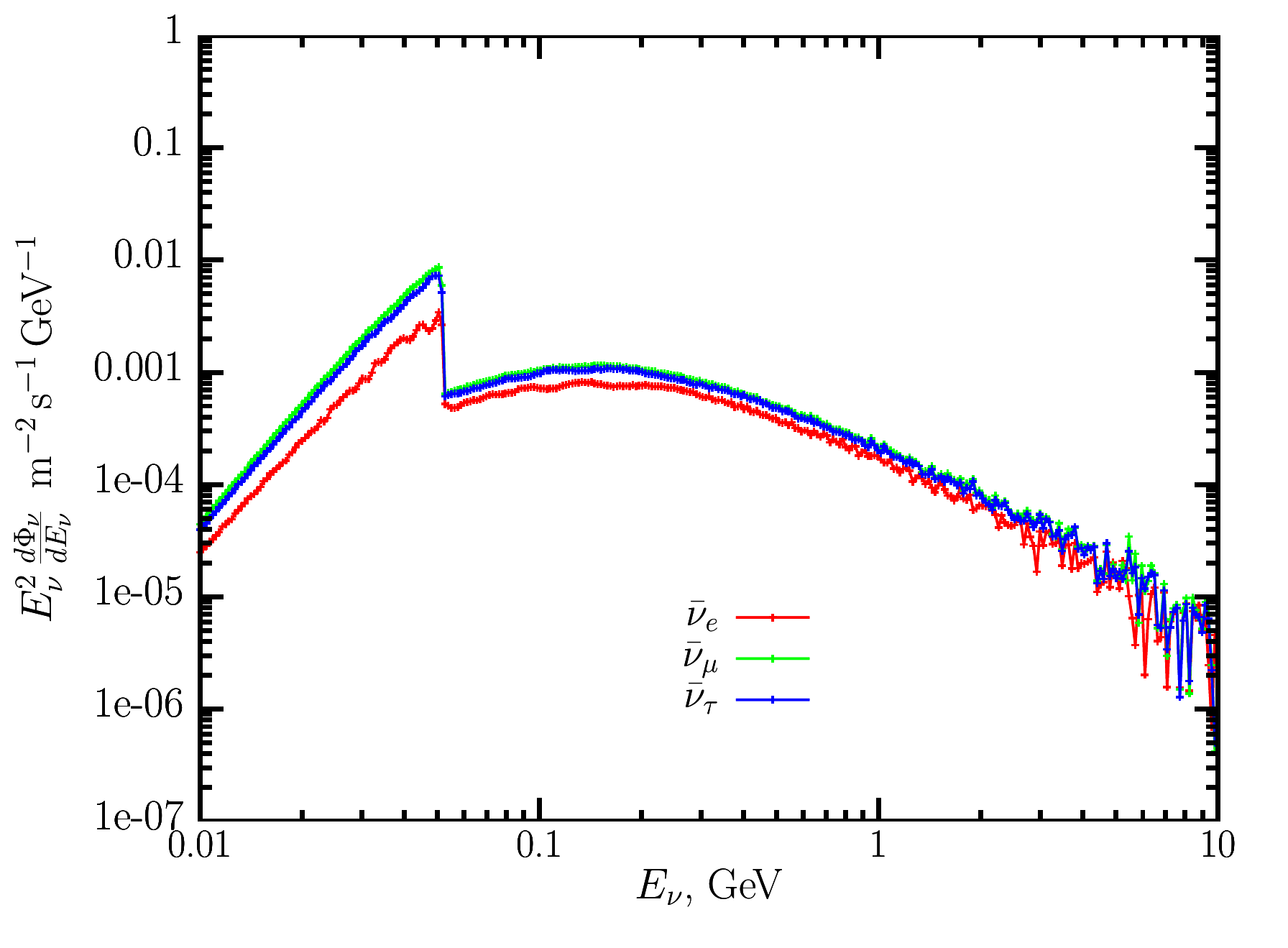}\\
      \includegraphics[width=9.0cm]{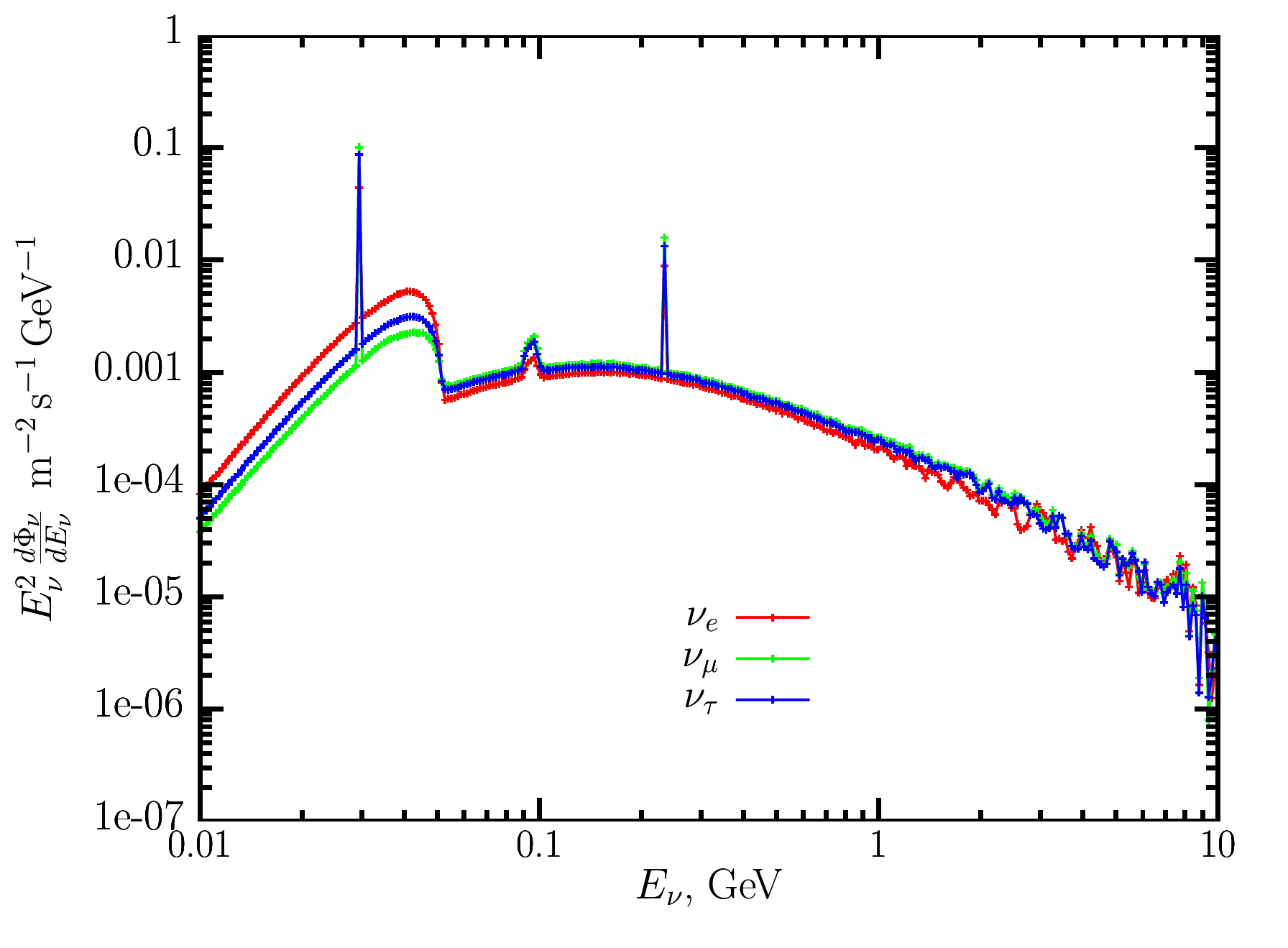} &
      \includegraphics[width=9.0cm]{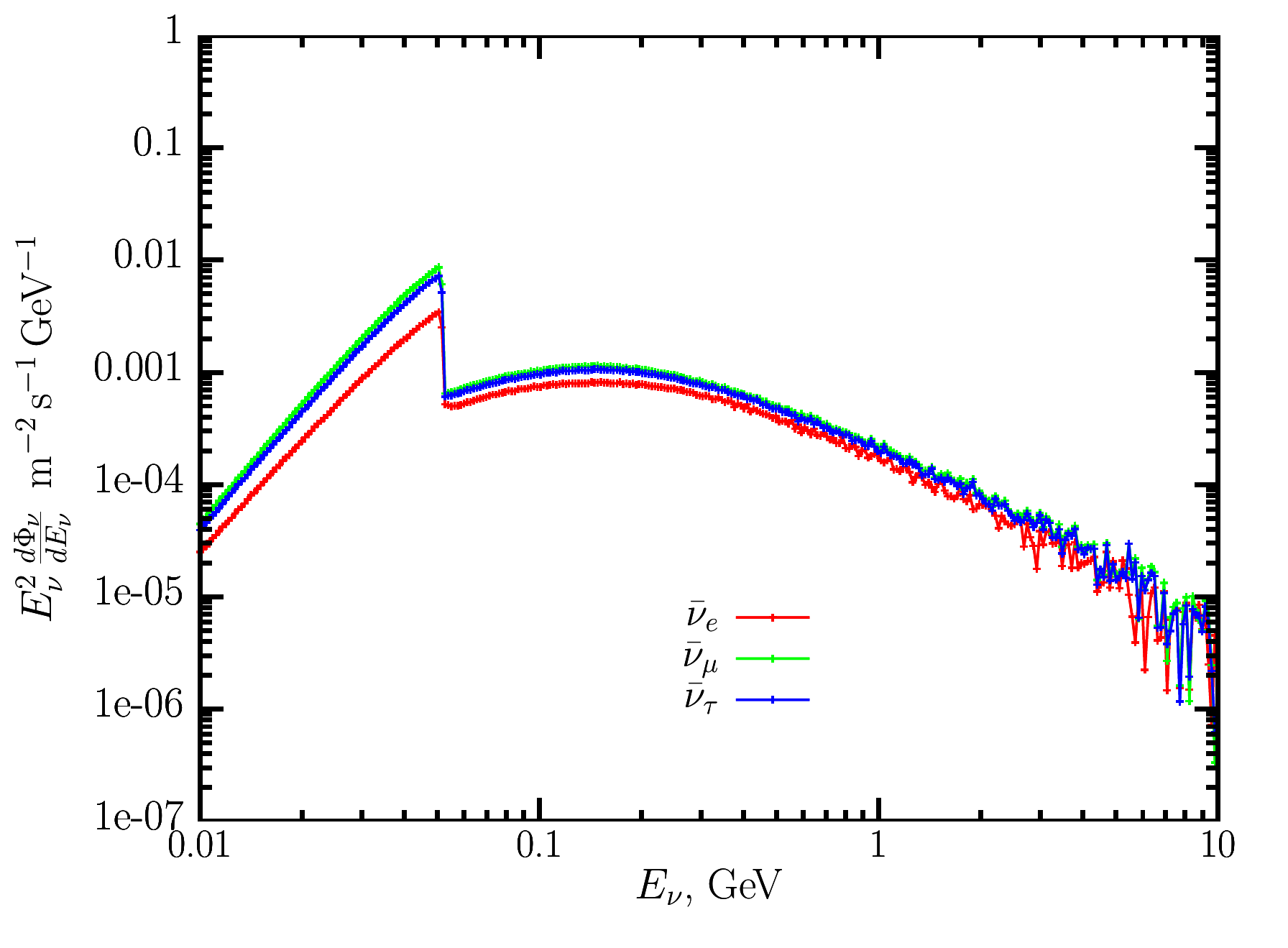}
    \end{tabular}
    \caption{\label{dFlux_osc} Fluxes of lunar neutrinos (left) and
      antineutrinos (right) calculated with account of neutrino
      oscillations for $\cos{\theta_M}>0$ (upper panels) and after averaging over position
      of the Moon on the sky (lower panels). }   
  \end{center}
\end{figure}

Impacts
from matter effect in the Moon and the Earth results in
deviations from averaged neutrino oscillation probabilities and are
clearly visible in those Figures. However, see Sec.\,\ref{conclude}, to make these studies
feasible, one needs to collect a sufficient amount of lunar neutrinos
{\it from each position of the Moon,} which implies  unrealistically large operation
time of experiment. In this case one may sum lunar neutrinos along the Moon
trajectories, which average the matter effect in the Earth, but keep
intact the matter effect in the Moon. In Fig.~\ref{dFlux_osc}
we show resulting spectra of lunar neutrinos and antineutrinos at the Earth
calculated for two cases: 1) assuming Moon to be above the horizon, which
implies no matter effect in the Earth; 2) averaging over position of the Moon
on the full sky. In realistic experimental setup additional averaging
of oscillations comes from finite energy and directional resolution. Detailed
study of this question is beyond the scope of this paper and
generically depends on the experimental parameters. We will comment
on that in the next Section. However, already from the plots of
Fig.\,\ref{dFlux_osc} one concludes that though the difference is
recognisable by eye, in a realistic experiment the
matter effect in the Earth has no observable effect at all.


Let us note in passing that although here we study the oscillations numerically,
one can use an analytical approach to describe evolution of low energy neutrinos, see
e.g.~\cite{Ioannisian:2004vv,Peres:2009xe}. Matter effect of neutrino
propagation in the Moon may be potentially used to obtain an
information about its inner structure similar to the studies for the
Earth, see
e.g. Refs.~\cite{Winter:2015zwx,Ioannisian:2017dkx,Bakhti:2020tcj}.

\section{Discussion and conclusions}
\label{conclude}
In the letter we calculated flux of neutrinos originated from cosmic
rays hitting the Moon surface. Apposed to the
earlier study~\cite{Miller:2006cn} we concentrated on low energy ($\lsim 
10$~GeV) part of neutrino energy spectrum. Although lunar neutrinos
have the same origin as atmospheric ones, absence of atmosphere at the Moon
makes spectrum of the former to be quite peculiar and sensitive to the
mass density of the surface layer. In particular, it 
appears to be shifted to lower energies where neutrinos originate from
decays of mesons stopped in the regolith.
Without neutrino oscillations the very low energy part (less than
about 52.8~MeV) of the lunar neutrino spectrum at the Earth exceeds the
atmospheric neutrino flux from the Moon's solid angle by up to an order
of magnitude for $\nu_e$ and $\bar{\nu}_\mu$ while the spectrum of $\nu_\mu$
exhibits two narrow peaks at energies $E_\nu\approx 29.8$~MeV
and 235.6~MeV corresponding to neutrinos from $\pi^+$ and $K^+$
decays. Oscillations of neutrinos mix those effects between all
neutrino flavours. At energies lower than about 20--50~MeV
  the lunar neutrino flux is comparable with that of DSNB within
  Moon's solid angle.
Contrary to atmospheric and diffuse neutrinos the lunar neutrino flux
is expected to endue a periodic 
dependence with amplitude about 12\% from small ellipticity of the Moon
orbit. These peculiarities, including explicit binding to the direction 
of the Moon, can be used to search for lunar neutrinos
in future neutrino experiments and to distinguish them from
atmospheric neutrino and diffuse supernova neutrino background.

At present, neutrino detectors used or planned for study of
sub-GeV neutrinos include water Cerenkov (WC) detectors such as
Hyper-Kamiokande~\cite{Abe:2011ts,Abe:2015zbg} and its predecessor
Super-Kamiokande, liquid scintillator (LS) and liquid argon time
projection chambers (LArTPC). The latter representative examples are
JUNO~\cite{An:2015jdp} and DUNE~\cite{Acciarri:2015uup},  
respectively. WC detectors can be more easily realized in large volume
which is crucial requirement for detection of the pretty low  neutrino flux
from the Moon. However, in neutrino-nucleon scattering correlation
between direction of charged lepton and incoming neutrino is weak for
neutrinos from sub-GeV energy range.
Moreover, for such neutrino energies an additional background in WC
detectors comes from invisible muons. LS and LArTPC detectors are free
of this type of the background. Still, typical detection channels used
in LS detectors do not provide very large information about neutrino
direction in the considered energy range. LArTPC detectors on the
other hand can provide not only energy but also a directional
information (see e.g.~\cite{Rott:2016mzs}) as for them it is possible
to reconstruct positions of charged lepton and
nucleon~~\cite{Anderson:2012vc}. Still, even if very large 
exposures can be realized here, at present an angular resolution of these
detectors is still quite poor as compared to the angular size of the Moon
on the sky. We conclude that detection of lunar neutrinos would require 
not large but huge neutrino detectors with (so far) exceptional energy
and,  especially, angular resolution -- the task which may one day
be feasible.


\section*{Acknowledgements}
We thank V. Izmodenov, A. Sanin, S. Troitsky for valuable discussions
and Yu. Kudenko, D. Naumov for clarification on the prospects of
ongoing and future neutrino experiments. 
This work is supported by the Ministry of science and higher education
   of the Russian Federation under the contract \# 075-15-2020-778 in the
   framework of the Large scientific projects program within the national
   project "Science".

\bibliographystyle{JHEP}
\bibliography{note}

\providecommand{\href}[2]{#2}\begingroup\raggedright\begin{thebibliography}{10}

\bibitem{Anchordoqui:2013dnh}
L.~A. Anchordoqui et~al., \emph{{Cosmic Neutrino Pevatrons: A Brand New Pathway
  to Astronomy, Astrophysics, and Particle Physics}},
  \href{https://doi.org/10.1016/j.jheap.2014.01.001}{\emph{JHEAp} {\bfseries
  1-2} (2014) 1--30}, [\href{https://arxiv.org/abs/1312.6587}{{\ttfamily
  1312.6587}}].

\bibitem{Vitagliano:2019yzm}
E.~Vitagliano, I.~Tamborra and G.~Raffelt, \emph{{Grand Unified Neutrino
  Spectrum at Earth: Sources and Spectral Components}},
  \href{https://doi.org/10.1103/RevModPhys.92.045006}{\emph{Rev. Mod. Phys.}
  {\bfseries 92} (2020) 045006},
  [\href{https://arxiv.org/abs/1910.11878}{{\ttfamily 1910.11878}}].

\bibitem{Miller:2006cn}
R.~Miller and T.~Cohen, \emph{{Cosmic-ray induced neutrino fluxes at the Moon:
  A semi-analytic approach}},
  \href{https://doi.org/10.1016/j.astropartphys.2006.03.009}{\emph{Astropart.
  Phys.} {\bfseries 25} (2006) 368--374}.

\bibitem{1965ICRC....2.1039V}
L.~V. {Volkova} and G.~T. {Zatsepin}, \emph{{The energy spectra of muons and
  neutrinos generated by cosmic rays in different substances.}},  in
  \emph{International Cosmic Ray Conference}, vol.~1 of \emph{International
  Cosmic Ray Conference}, p.~1039, Jan., 1965.

\bibitem{Volkova1989}
L.~V. Volkova, \emph{Fluxes of Muons and Neutrinos Generated by Primary
  Radiation on the Moon}, pp.~141--144.
\newblock Springer Netherlands, Dordrecht, 1989.

\bibitem{doi:10.1002/2016JA023308}
Y.~Li, X.~Zhang, W.~Dong, Z.~Ren, T.~Dong and A.~Xu, \emph{Simulation of the
  production rates of cosmogenic nuclides on the moon based on geant4},
  \href{https://doi.org/10.1002/2016JA023308}{\emph{Journal of Geophysical
  Research: Space Physics} {\bfseries 122} (2017) 1473--1486}.

\bibitem{Moskalenko:2007mk}
I.~V. Moskalenko and T.~A. Porter, \emph{{The Gamma-ray Albedo of the Moon}},
  \href{https://doi.org/10.1086/522828}{\emph{Astrophys. J.} {\bfseries 670}
  (2007) 1467--1472}, [\href{https://arxiv.org/abs/0708.2742}{{\ttfamily
  0708.2742}}].

\bibitem{Abdo:2012nfa}
A.~Abdo et~al., \emph{{FERMI observations of $\gamma$-ray emission from the
  Moon}}, \href{https://doi.org/10.1088/0004-637X/758/2/140}{\emph{Astrophys.
  J.} {\bfseries 758} (2012) 140}.

\bibitem{Cerutti:2016gts}
{\scshape Fermi-LAT} collaboration, M.~Ackermann et~al., \emph{{Measurement of
  the high-energy gamma-ray emission from the Moon with the Fermi Large Area
  Telescope}}, \href{https://doi.org/10.1103/PhysRevD.93.082001}{\emph{Phys.
  Rev. D} {\bfseries 93} (2016) 082001},
  [\href{https://arxiv.org/abs/1604.03349}{{\ttfamily 1604.03349}}].

\bibitem{Distefano:2011zza}
{\scshape ANTARES} collaboration, C.~Distefano, \emph{{On the detection of the
  Moon shadow with the ANTARES neutrino telescope}},
  \href{https://doi.org/10.1016/j.nima.2010.06.228}{\emph{Nucl. Instrum. Meth.
  A} {\bfseries 626-627} (2011) S223--S225}.

\bibitem{Albert:2018yoj}
{\scshape ANTARES} collaboration, A.~Albert et~al., \emph{{The cosmic ray
  shadow of the Moon observed with the ANTARES neutrino telescope}},
  \href{https://doi.org/10.1140/epjc/s10052-018-6451-3}{\emph{Eur. Phys. J. C}
  {\bfseries 78} (2018) 1006},
  [\href{https://arxiv.org/abs/1807.11815}{{\ttfamily 1807.11815}}].

\bibitem{Boersma:2010zz}
{\scshape IceCube} collaboration, D.~Boersma, L.~Gladstone and A.~Karle,
  \emph{{Moon Shadow Observation by IceCube}},
  \href{https://arxiv.org/abs/1002.4900}{{\ttfamily 1002.4900}}.

\bibitem{Aartsen:2013zka}
{\scshape IceCube} collaboration, M.~Aartsen et~al., \emph{{Observation of the
  cosmic-ray shadow of the Moon with IceCube}},
  \href{https://doi.org/10.1103/PhysRevD.89.102004}{\emph{Phys. Rev. D}
  {\bfseries 89} (2014) 102004},
  [\href{https://arxiv.org/abs/1305.6811}{{\ttfamily 1305.6811}}].

\bibitem{Fargion:2017lok}
D.~Fargion, P.~Oliva, P.~G. de~Sanctis~Lucentini and M.~Y. Khlopov,
  \emph{{Signals of HE atmospheric $\mu$ decay in flight around the Sun's
  albedo versus astrophysical $\nu_\mu$ and $\nu_\tau$ traces in the Moon
  shadow}}, \href{https://doi.org/10.1142/S021827181841002X}{\emph{Int. J. Mod.
  Phys. D} {\bfseries 27} (2018) 1841002},
  [\href{https://arxiv.org/abs/1706.09352}{{\ttfamily 1706.09352}}].

\bibitem{Agostinelli:2002hh}
{\scshape GEANT4} collaboration, S.~Agostinelli et~al., \emph{{GEANT4--a
  simulation toolkit}},
  \href{https://doi.org/10.1016/S0168-9002(03)01368-8}{\emph{Nucl. Instrum.
  Meth. A} {\bfseries 506} (2003) 250--303}.

\bibitem{Tanabashi:2018oca}
{\scshape Particle Data Group} collaboration, M.~Tanabashi et~al.,
  \emph{{Review of Particle Physics}},
  \href{https://doi.org/10.1103/PhysRevD.98.030001}{\emph{Phys. Rev. D}
  {\bfseries 98} (2018) 030001}.

\bibitem{Moon}
S.~R. Taylor, \emph{{Lunar science: a post-Apollo view; scientific results and
  insights from the lunar samples.}}, {\emph{New York: Pergamon Press} (1975)
  }.

\bibitem{Ponomarev:1973ya}
L.~Ponomarev, \emph{{Molecular structure effects on atomic and nuclear capture
  of mesons}},
  \href{https://doi.org/10.1146/annurev.ns.23.120173.002143}{\emph{Ann. Rev.
  Nucl. Part. Sci.} {\bfseries 23} (1973) 395--430}.

\bibitem{Measday:2001yr}
D.~Measday, \emph{{The nuclear physics of muon capture}},
  \href{https://doi.org/10.1016/S0370-1573(01)00012-6}{\emph{Phys. Rept.}
  {\bfseries 354} (2001) 243--409}.

\bibitem{Spitz:2014hwa}
J.~Spitz, \emph{{Cross Section Measurements with Monoenergetic Muon
  Neutrinos}}, \href{https://doi.org/10.1103/PhysRevD.89.073007}{\emph{Phys.
  Rev. D} {\bfseries 89} (2014) 073007},
  [\href{https://arxiv.org/abs/1402.2284}{{\ttfamily 1402.2284}}].

\bibitem{Rott:2012qb}
C.~Rott, J.~Siegal-Gaskins and J.~F. Beacom, \emph{{New Sensitivity to Solar
  WIMP Annihilation using Low-Energy Neutrinos}},
  \href{https://doi.org/10.1103/PhysRevD.88.055005}{\emph{Phys. Rev. D}
  {\bfseries 88} (2013) 055005},
  [\href{https://arxiv.org/abs/1208.0827}{{\ttfamily 1208.0827}}].

\bibitem{Bernal:2012qh}
N.~Bernal, J.~Martin-Albo and S.~Palomares-Ruiz, \emph{{A novel way of
  constraining WIMPs annihilations in the Sun: MeV neutrinos}},
  \href{https://doi.org/10.1088/1475-7516/2013/08/011}{\emph{JCAP} {\bfseries
  08} (2013) 011}, [\href{https://arxiv.org/abs/1208.0834}{{\ttfamily
  1208.0834}}].

\bibitem{Rott:2015nma}
C.~Rott, S.~In, J.~Kumar and D.~Yaylali, \emph{{Dark Matter Searches for
  Monoenergetic Neutrinos Arising from Stopped Meson Decay in the Sun}},
  \href{https://doi.org/10.1088/1475-7516/2015/11/039}{\emph{JCAP} {\bfseries
  11} (2015) 039}, [\href{https://arxiv.org/abs/1510.00170}{{\ttfamily
  1510.00170}}].

\bibitem{Rott:2016mzs}
C.~Rott, S.~In, J.~Kumar and D.~Yaylali, \emph{{Directional Searches at DUNE
  for Sub-GeV Monoenergetic Neutrinos Arising from Dark Matter Annihilation in
  the Sun}}, \href{https://doi.org/10.1088/1475-7516/2017/01/016}{\emph{JCAP}
  {\bfseries 01} (2017) 016},
  [\href{https://arxiv.org/abs/1609.04876}{{\ttfamily 1609.04876}}].

\bibitem{Battistoni:2005pd}
G.~Battistoni, A.~Ferrari, T.~Montaruli and P.~Sala, \emph{{The atmospheric
  neutrino flux below 100-MeV: The FLUKA results}},
  \href{https://doi.org/10.1016/j.astropartphys.2005.03.006}{\emph{Astropart.
  Phys.} {\bfseries 23} (2005) 526--534}.

\bibitem{Honda:2015fha}
M.~Honda, M.~Sajjad~Athar, T.~Kajita, K.~Kasahara and S.~Midorikawa,
  \emph{{Atmospheric neutrino flux calculation using the NRLMSISE-00
  atmospheric model}},
  \href{https://doi.org/10.1103/PhysRevD.92.023004}{\emph{Phys. Rev. D}
  {\bfseries 92} (2015) 023004},
  [\href{https://arxiv.org/abs/1502.03916}{{\ttfamily 1502.03916}}].

\bibitem{OHare:2015utx}
C.~A.~J. O'Hare, A.~M. Green, J.~Billard, E.~Figueroa-Feliciano and L.~E.
  Strigari, \emph{{Readout strategies for directional dark matter detection
  beyond the neutrino background}},
  \href{https://doi.org/10.1103/PhysRevD.92.063518}{\emph{Phys. Rev. D}
  {\bfseries 92} (2015) 063518},
  [\href{https://arxiv.org/abs/1505.08061}{{\ttfamily 1505.08061}}].

\bibitem{https://doi.org/10.1002/2013JE004559}
J.~G. Williams, A.~S. Konopliv, D.~H. Boggs, R.~S. Park, D.-N. Yuan, F.~G.
  Lemoine et~al., \emph{Lunar interior properties from the grail mission},
  \href{https://doi.org/https://doi.org/10.1002/2013JE004559}{\emph{Journal of
  Geophysical Research: Planets} {\bfseries 119} (2014) 1546--1578}.

\bibitem{DZIEWONSKI1981297}
A.~M. Dziewonski and D.~L. Anderson, \emph{Preliminary reference earth model},
  \href{https://doi.org/https://doi.org/10.1016/0031-9201(81)90046-7}{\emph{Physics
  of the Earth and Planetary Interiors} {\bfseries 25} (1981) 297--356}.

\bibitem{Esteban:2018azc}
I.~Esteban, M.~Gonzalez-Garcia, A.~Hernandez-Cabezudo, M.~Maltoni and
  T.~Schwetz, \emph{{Global analysis of three-flavour neutrino oscillations:
  synergies and tensions in the determination of $\theta_{23}$, $\delta_{CP}$,
  and the mass ordering}},
  \href{https://doi.org/10.1007/JHEP01(2019)106}{\emph{JHEP} {\bfseries 01}
  (2019) 106}, [\href{https://arxiv.org/abs/1811.05487}{{\ttfamily
  1811.05487}}].

\bibitem{Ioannisian:2004vv}
A.~N. Ioannisian, N.~A. Kazarian, A.~Y. Smirnov and D.~Wyler, \emph{{A Precise
  analytical description of the earth matter effect on oscillations of low
  energy neutrinos}},
  \href{https://doi.org/10.1103/PhysRevD.71.033006}{\emph{Phys. Rev. D}
  {\bfseries 71} (2005) 033006},
  [\href{https://arxiv.org/abs/hep-ph/0407138}{{\ttfamily hep-ph/0407138}}].

\bibitem{Peres:2009xe}
O.~L.~G. Peres and A.~Y. Smirnov, \emph{{Oscillations of very low energy
  atmospheric neutrinos}},
  \href{https://doi.org/10.1103/PhysRevD.79.113002}{\emph{Phys. Rev. D}
  {\bfseries 79} (2009) 113002},
  [\href{https://arxiv.org/abs/0903.5323}{{\ttfamily 0903.5323}}].

\bibitem{Winter:2015zwx}
W.~Winter, \emph{{Atmospheric Neutrino Oscillations for Earth Tomography}},
  \href{https://doi.org/10.1016/j.nuclphysb.2016.03.033}{\emph{Nucl. Phys. B}
  {\bfseries 908} (2016) 250--267},
  [\href{https://arxiv.org/abs/1511.05154}{{\ttfamily 1511.05154}}].

\bibitem{Ioannisian:2017dkx}
A.~Ioannisian, A.~Smirnov and D.~Wyler, \emph{{Scanning the Earth with solar
  neutrinos and DUNE}},
  \href{https://doi.org/10.1103/PhysRevD.96.036005}{\emph{Phys. Rev. D}
  {\bfseries 96} (2017) 036005},
  [\href{https://arxiv.org/abs/1702.06097}{{\ttfamily 1702.06097}}].

\bibitem{Bakhti:2020tcj}
P.~Bakhti and A.~Y. Smirnov, \emph{{Oscillation tomography of the Earth with
  solar neutrinos and future experiments}},
  \href{https://doi.org/10.1103/PhysRevD.101.123031}{\emph{Phys. Rev. D}
  {\bfseries 101} (2020) 123031},
  [\href{https://arxiv.org/abs/2001.08030}{{\ttfamily 2001.08030}}].

\bibitem{Abe:2011ts}
K.~Abe et~al., \emph{{Letter of Intent: The Hyper-Kamiokande Experiment ---
  Detector Design and Physics Potential ---}},
  \href{https://arxiv.org/abs/1109.3262}{{\ttfamily 1109.3262}}.

\bibitem{Abe:2015zbg}
{\scshape Hyper-Kamiokande Proto-} collaboration, K.~Abe et~al., \emph{{Physics
  potential of a long-baseline neutrino oscillation experiment using a J-PARC
  neutrino beam and Hyper-Kamiokande}},
  \href{https://doi.org/10.1093/ptep/ptv061}{\emph{PTEP} {\bfseries 2015}
  (2015) 053C02}, [\href{https://arxiv.org/abs/1502.05199}{{\ttfamily
  1502.05199}}].

\bibitem{An:2015jdp}
{\scshape JUNO} collaboration, F.~An et~al., \emph{{Neutrino Physics with
  JUNO}}, \href{https://doi.org/10.1088/0954-3899/43/3/030401}{\emph{J. Phys.
  G} {\bfseries 43} (2016) 030401},
  [\href{https://arxiv.org/abs/1507.05613}{{\ttfamily 1507.05613}}].

\bibitem{Acciarri:2015uup}
{\scshape DUNE} collaboration, R.~Acciarri et~al., \emph{{Long-Baseline
  Neutrino Facility (LBNF) and Deep Underground Neutrino Experiment (DUNE)}:
  {Conceptual Design Report, Volume 2: The Physics Program for DUNE at LBNF}},
  \href{https://arxiv.org/abs/1512.06148}{{\ttfamily 1512.06148}}.

\bibitem{Anderson:2012vc}
C.~Anderson et~al., \emph{{The ArgoNeuT Detector in the NuMI Low-Energy beam
  line at Fermilab}},
  \href{https://doi.org/10.1088/1748-0221/7/10/P10019}{\emph{JINST} {\bfseries
  7} (2012) P10019}, [\href{https://arxiv.org/abs/1205.6747}{{\ttfamily
  1205.6747}}].

\end{thebibliography}\endgroup

\end{document}